\documentclass[namedreferences]{solarphysics}

\usepackage[optionalrh]{spr-sola-addons} 
\usepackage{graphicx}        
\usepackage{amssymb}    
\usepackage{longtable}
\usepackage{rotating}
\usepackage{booktabs}
\usepackage{color}           
\graphicspath{{./}{figures/}}
\usepackage{hyperref}
\hypersetup{
    colorlinks=true,
    linkcolor=red,
    citecolor = blue,
    filecolor=cyan,      
    urlcolor=magenta,
}

\newcommand{\aap}{    {\it Astron. Astrophys.}}

\newcommand{\apj}{    {\it Astrophys. J.}}
\newcommand{\apjl}{   {\it Astrophys. J. Lett.}}

\newcommand{\grl}{    {\it Geophys. Res. Lett.}}

\newcommand{\jgr}{    {\it J. Geophys. Res.}}

\newcommand{\solphys}{{\it Solar Phys.}}
 
\newcommand{\ssr}{    {\it Space Sci. Rev.}} 

\chardef\us=`\_

\begin{document}

\begin{article}
\begin{opening}

\title{Imaging and Spectral Observations of a Type-II Radio Burst Revealing the Section of the CME-Driven Shock that Accelerates Electrons}
\author[addressref={aff1},corref,email={satabdwa.m@iiap.res.in}]{\inits{S.}\fnm{Satabdwa}~\lnm{Majumdar}\orcid{0000-0002-6553-3807}}
\author[addressref=aff2]{\inits{S.P.}\fnm{Srikar~Paavan}~\lnm{Tadepalli}}
\author[addressref={aff1,aff3}]{\inits{S.S.}\fnm{Samriddhi~Sankar}~\lnm{Maity}}
\author[addressref={aff1,aff4}]{\inits{K.}\fnm{Ketaki}~\lnm{Deshpande}\orcid{0000-0001-6861-6328}}
\author[addressref={aff1,aff5}]{\inits{A.}\fnm{Anshu}~\lnm{Kumari}\orcid{0000-0001-5742-9033}}
\author[addressref={aff1,aff6}]{\inits{R.}\fnm{Ritesh}~\lnm{Patel}\orcid{0000-0001-8504-2725}}
\author[addressref={aff7}]{\inits{N.}\fnm{Nat}~\lnm{Gopalswamy}\orcid{0000-0001-5894-9954}}

\address[id=aff1]{Indian Institute of Astrophysics, Koramangala 2nd Block, Bangalore, Karnataka - 560034, India }
\address[id=aff2]{ISRO UR Rao Satellite Center, Vimanapura, Bangalore, Karnataka - 560017, India }
\address[id=aff3]{Indian Institute of Science, CV Raman Rd, Bangalore, Karnataka - 560012, India}
\address[id=aff4]{Sir Parashurambhau College, Pune, Maharashtra - 411030, India}
\address[id=aff5]{Department of Physics, University of Helsinki, P.O. Box 64, 00014 FI Helsinki, Finland}
\address[id=aff6]{Aryabhatta Research Institute of Observational Sciences, Beluwakhan, Uttarakhand - 263002, India}
\address[id=aff7]{NASA Goddard Space Flight Center, Code 671, Greenbelt, 20771 MD, USA}
\runningauthor{M.Satabdwa et al.}
\runningtitle{Multiwavelength Study of a CME on 26 January 2014}

\begin{abstract}

We report on a multi-wavelength analysis of the 26 January 2014 solar eruption involving a coronal mass ejection (CME) and a Type-II radio burst, performed by combining data from various space-and ground-based instruments. An increasing standoff distance with height shows the presence of a strong shock, which further manifests itself in the continuation of the metric Type-II burst into the decameter-hectometric (DH) domain. A plot of speed versus position angle (PA) shows different points on the CME leading edge travelled with different speeds. From the starting frequency of the Type-II burst and white-light data, we find that the shock signature producing the Type-II burst might be coming from the flanks of the CME. Measuring the speeds of the CME flanks, we find the southern flank to be at a higher speed than the northern flank; further the radio contours from Type-II imaging data showed that the burst source was coming from the southern flank of the CME. From the standoff distance at the CME nose, we find that the local Alf\'{v}en speed is close to the white-light shock speed, thus causing the Mach number to be small there. Also, the presence of a streamer near the southern flank appears to have provided additional favorable conditions for the generation of shock-associated radio emission. 
These results provide conclusive evidence that the Type-II emission could originate from the flanks of the CME, which in our study is from the the southern flank of the CME.

\end{abstract}
\keywords{Activity, corona, coronal mass ejections (CMEs), flares,  radio radiation, radio bursts}
\end{opening}

\section{Introduction}
     \label{S-Introduction} 
     
Coronal mass ejections (CMEs) are some of the most energetic explosions happening in the solar atmosphere, expelling large amounts of plasma and magnetic field into the heliosphere with speeds ranging from a few tens to a few thousands of \,km s$^{-1}$, see \cite{article}, for a review. CMEs are also one of the major drivers of space weather, which can drastically affect human technological systems  \citep{1991JGR....96.7831G,1993JGR....9818937G, 2012JSWSC...2A..01R, 2013AnGeo..31.1251K}. CMEs can also drive magneto-hydrodynamic shocks that can accelerate energetic particles,a key aspect in the understanding of space weather \citep[]{1978SoPh...57..429K, 1999SSRv...90..413R, 2020ApJ...900...75K}. In order to understand the shock-driving capability of CMEs, we need to investigate early CME kinematics near the Sun. CMEs leave imprints in different wavelengths, so it is necessary to stitch multiwavelength information for a better understanding of CME behaviour \citep{2006SSRv..123..341P, Vrsnak2006, Zucca2014, kumari2017b, morosan2020electron}.\\

CMEs show a wide range in their kinematic properties \citep[and references therein]{2004JGRA..109.7105Y}, with a three-phase kinematic profile, an initial gradual rise phase, an impulsive acceleration phase, and then a final phase with constant or decreasing speed \citep{Zhang2001, Zhang_2004, article}. A major concern here is that for measurements done on the plane of the sky, the results may suffer from projection effects \citep{2018ApJ...863...57B}. In this regard, to reduce such projection effects, \cite{2009SoPh..256..111T} developed the Graduated Cylindrical Shell (GCS) model which uses forward modeling to fit a flux-rope to coronagraph images taken from multiple vantage points, and hence to reconstruct the 3D structure of the CME. CMEs are also capable of driving shocks in the low corona and interplanetary (IP) medium \citep{1987sowi.conf..181H}. Often shocks associated with CMEs are observed in white light images, and in such cases, the shock signatures can be tracked directly from the white-light CME images \citep{2000JGR...105.5081S,vourlidas2008}. In cases where there is an associated Type-II burst, the Type-II burst can be used to track the shock signature \citep{2006GMS...165..207G, 2010ApJ...712..188R, kumari2017a, Kahler2019}. \cite{2009SoPh..259..227G} reported on the relationship between Type-II bursts and CMEs showing their combined evolution from the corona into the IP medium. The authors reported that Type-II bursts provide the earliest signature of a shock that forms within a fraction of solar radius above the surface \citep{cane1984type, cho2013high, 2013AdSpR..51.1981G, kumari2017a}.\\

\cite{1983ApJ...267..837H} suggested that the electrons that are responsible for the Type-II burst might get accelerated from the shock flanks, which implies that the height inferred from the Type-II burst location might be smaller than the height of the CME leading edge at that particular time. Also, given that the frequency drift rate of Type-II bursts is related to the speed of the shock and the density scale height of the ambient corona \citep{2009SoPh..259..227G}, it is important to understand which section of the shock surface produces the Type-II burst source. In this regard, \citet{2020A&A...639A..56J} used a radio triangulation technique to understand the origin of two consecutive Type-II bursts. Metric Type-II bursts are also sometimes found to show their fundamental and harmonic emission band being split into two parallel lanes. \cite{1974IAUS...57..389S} proposed an explanation for this band splitting in terms of the emission coming from the upstream and downstream shock regions, with the observational support to the theory reported by \cite{vrsnak_2001}. Band splitting in Type-II bursts has proven to be useful to estimate the ambient coronal magnetic field in the inner corona  \citep{1974IAUS...57..389S,vrsnak_2001,vrsnak_2002,vrsnak_2004,2007ApJ...665..799C, kumari2017c}. 
Direct estimate of the magnetic field in the inner corona from a Type-II burst using an empirical electron-density distribution was also reported by \cite{kumari2019}. Further, the ambient coronal magnetic-field strength beyond the inner corona can be estimated by the shock stand-off distance and radius of curvature of the flux-rope cross-section as described by \cite{2011ApJ...736L..17G}. Thus for a better understanding of the kinematics of CMEs and its interaction with the ambient medium we need to consider all aspects of the Type-II bursts and the associated CMEs. \\

We use the Type-II -- CME connection to understand several aspects of the evolution of a CME, as it propagates into the heliosphere. We analyze the 26 January 2014 CME by combining white-light, radio, extreme ultra-violet (EUV), and X-ray data from various space-and ground-based instruments. We outline the data sources that we have used in Section \ref{data} and provide a brief description and timeline of the event in Section \ref{description}, followed by our results in Section \ref{results}, and conclusions in Section \ref{summary}.

\section{Data Selection} \label{data} 
We have used white-light coronagraph data from the \textit{ Large Angle Spectrometric Coronagraph} C2 and C3 (LASCO: \citet{Brueckner95}) onboard the \textit{Solar and Heliospheric Observatory} (SOHO), COR-1 and COR-2 coronagraphs of the \textit{Sun Earth Connection Coronal and Heliospheric Investigation} (SECCHI: \cite{2008SSRv..136...67H}) package on the \textit{Solar Terrestrial Relations Observatory} (STEREO: \cite{2008SSRv..136....5K}). We have used EUV data from different passbands of the \textit{Atmospheric Imaging Assembly} (AIA: \cite{aia}) onboard \textit{Solar Dynamics Observatory} (SDO), \textit{Extreme Ultra Violet Imager} (EUVI) 
onboard STEREO and \textit{Extreme-ultraviolet Imaging Telescope} (EIT: \cite{SOHOEIT}) onboard SOHO. X-ray flux from the \textit{Geostationary Operational Environmental Satellite} (GOES) (1--8 \,\AA~ channel) provides the flare context. We have used radio data from the \textit{Gauribidanur Low-frequency Solar Spectrograph} (GLOSS: \cite{2014SoPh..289.3995K}),  the \textit{Gauribidanur RAdio heliograPH} (GRAPH; \citep{1998SoPh..181..439R}), Learmonth station of the \textit{Radio Solar Telescope Network} (RSTN), \textit{Compound Astronomical Low frequency Low cost Instrument for Spectroscopy and Transportable Observatory} (e-CALLISTO: \cite{2013EGUGA..15.2027M}) at the Gauribidanur Radio Observatory and the SWAVES instrument \citep{SWAVES} onboard STEREO. Data from the Coordinated Data Analysis Workshop (CDAW: \citet{Gopalswamy2009EM&PG}) catalogue which lists the properties of CMEs detected manually in LASCO images onboard SOHO are also used.

\section{Event Description and Timeline}\label{description}

\begin{figure}
    \centering
    \includegraphics[width=0.45\textwidth,clip=]{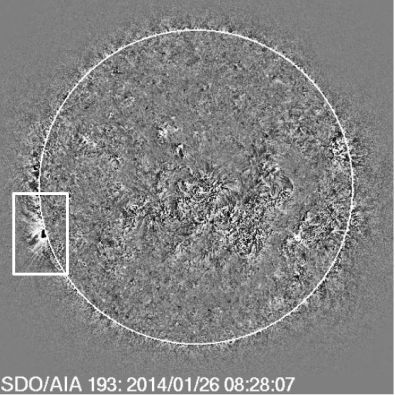}
    \includegraphics[width=0.45\textwidth,clip=]{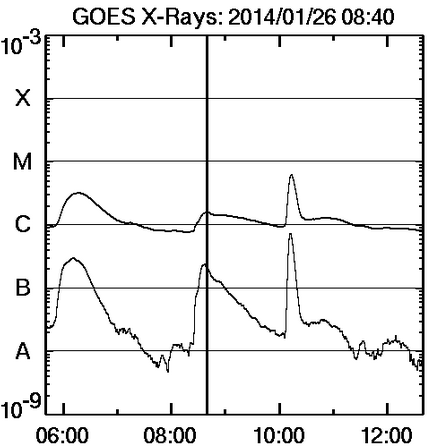}
    \caption{{\it Left:} The source region of the CME as observed in SDO/AIA 193 \,\AA~ shown in the enclosed rectangle at the eastern limb. The image is a running : difference image made at 08:28:07 UT. The white circle represents the solar disk.
    {\it Right:} The GOES soft X-ray light curve of the associated C1.5 class flare on 26 January 2014. Note that the flare is occulted, so the actual soft X-ray intensity is expected to be higher than that shown. }
    \label{flare}
\end{figure}

\begin{figure}[!ht]
    \centering
    \centerline{\hspace*{0.05\textwidth}
     \includegraphics[width=0.33\textwidth,clip=]{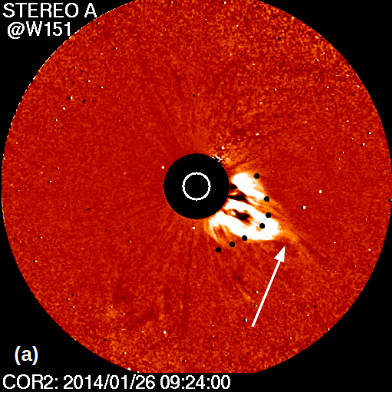}
     \includegraphics[width=0.33\textwidth,clip=]{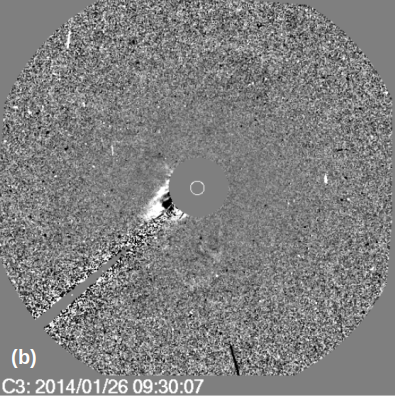}
     \includegraphics[width=0.33\textwidth,clip=]{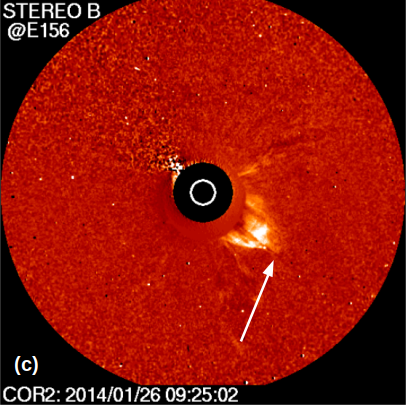}
     }
    \caption{The CME on 26 January, 2014 observed with (a) STEREO/COR-2A; (b) SOHO/LASCO-C3; and (c) STEREO/COR-2B. The diffused shock structure beyond the bright flux-rope (observed in a very limited area beyond the flux-rope) can be seen in STEREO A and STEREO B views as marked with white arrow. The location of STEREO-A and STEREO-B were W151 and E156, respectively. The representative position angles for which the speed is measured in Figure~\ref{fr_shock}d are shown in black dots in (a) 
    }
    \label{cme_views}
\end{figure}

We present a description of the event at the time of eruption in this section. The CME on 26 January 2014 was associated with a C1.5 flare from NOAA Active Region 11967 located at S16E106, which is 16$^{\circ}$ behind the east limb (see Figure \ref{flare}). Since the flare is partly occulted, it is likely that the flare size is underestimated. The flare starts at $\approx$ 8:24 UT, peaks at $\approx$ 8:36 UT and ends at $\approx$ 9:48 UT (see Figure \ref{flare}, right panel). The CME is observed fully by LASCO-C2 and-C3) and SECCHI (COR-1, COR-2) on STEREO-A and-B.  This CME is listed in the CDAW catalogue with a first appearance time of 8:36:05 UT in the LASCO-C2 field of view (FOV) with an average speed of 1088 \,km s$^{-1}$. The partial halo CME is propagating in the southeast direction with a central position angle (CPA) of 125$^{\circ}$ in the LASCO FOV. The CME decelerates in the LASCO-C2 and COR-1, COR-2 FOV, indicating that the initial acceleration ended before the CME appeared in the FOV of COR-1 and LASCO-C2. The source region of the CME is also identified in the EUV images taken at 193 \,\AA~ by SDO/AIA. Figure \ref{flare} shows the location of the source region as observed by SDO/AIA at 193 \,\AA~ and the associated GOES soft X-ray profile. At the time of the CME eruption, STEREO-A and-B were located at W151 and E156, respectively. Therefore the CME is a limb event (W103) in the STEREO-A FOV and a disk event (W50) in the STEREO-B FOV (the appearance of the CME in COR-2A, LASCO C3, and COR-2B FOV is shown in Figure~\ref{cme_views}) implying that STEREO-A and LASCO measurements would not suffer much from projection effects. \\

\begin{figure}
    \centering
  \includegraphics[width=1\textwidth]{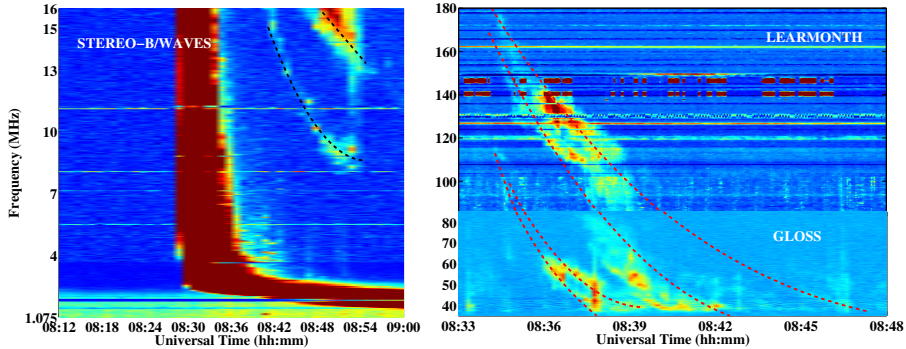}
    \caption{ The dynamic spectra of DH continuation of the metric Type-II burst (left panel) as observed by STEREO-B/WAVES (space-based) and the metric Type-II burst (right panel) recorded with the GLOSS and Learmonth spectrographs (ground-based). In the dynamic spectra recorded by the Learmonth spectrograph the start frequency of the Type-II bursts is $\approx$ 115 \,MHz. The Type-II burst shows fundamental-harmonic (FH) structures in metric-DH wavelengths. 
    The band splitting is seen in both FH bands. The FH bands are indicated with black dotted lines (left panel). The split bands are shown in red dotted lines (right panel). 
    }
    \label{typeII}
\end{figure}

 The dynamic spectra of the decameter--hectometric (DH) Type-II from STEREO-B/WAVES is shown in Figure \ref{typeII} (left panel). The dynamic spectra of the metric Type-II radio burst recorded with the Learmonth spectrograph belonging to the \textit{Radio Solar Telescope Network} (RSTN) and the GLOSS are combined and shown in Figure \ref{typeII} (right panel). The Type-II burst has fundamental-harmonic structure in the metric and DH domains. The starting frequency of the fundamental component is $\approx$ 115 \,MHz at $\approx$ 8:34 UT as seen in the Learmonth spectrograph. The burst then drifted towards the lower frequencies and continued to $\approx$ 7.46 MHz in the DH domain at $\approx$ 8:52 UT. The Type-II emission shows band-splitting. The continuation of the Type-II emission from metric to DH domain indicates the presence of a strong shock that is capable of propagating long distances, thus producing Type-II emission in different spectral domains \cite[also see][]{1999GeoRL..26.1573B, 2005JGRA..11012S07G,2009SoPh..259..227G}. It is important to note that the DH Type-II burst was observed only in STEREO, not in Wind/WAVES because the source region was behind the limb. Also it is best observed in STEREO-B because it observes more shock surface. STEREO-A sees only the harmonic. We also note that the height of formation of the Type-II associated shock wave ($\approx~1.3$ R$_{\odot}$) was below the COR-1 FOV, which did not enable us to determine the CME kinematics at the onset of the Type-II burst accurately. We summarise the timeline of the event as observed by the different instruments in Table~\ref{table}.

\begin{sidewaystable}
\centering
 {\setlength\doublerulesep{1.4pt}   
  \begin{tabular}{cccc}
  \toprule[0.8pt]\midrule[0.3pt]
      Phenomenon & Data Source & Parameter & Observations  \\
\hline
 Type-II burst & Learmonth spectrograph & Starting frequency (time) & 115 MHz (8:34 UT) \\
    Type-II burst & Gauribidanur spectrograph & Ending frequency (time) & 35 MHz (8:40 UT) \\
     Type-II burst & STEREO-B/WAVES & Starting frequency (time) & 16.46 MHz (8:40 UT) \\
    Type-II burst & STEREO-B/WAVES & Ending frequency (time) & 7.78 MHz (8:51 UT) \\
    \hline
    Flare & GOES X-Rays & Starting time & 8:24 UT \\
    Flare & GOES X-Rays & Peaking time & 8:36 UT \\
    Flare & GOES X-Rays & Ending time & 9:48 UT \\
    \hline
    CME & LASCO-C2 & First appearance height (time) & 2.27 R$_{\odot}$ (8:36 UT) \\
    CME & LASCO-C3 & First appearance height (time) & 6.01 R$_{\odot}$ (9:06 UT) \\
    CME & STEREO-A/COR 1 & First appearance height (time) & 3.14 R$_{\odot}$ (8:45 UT) \\
    CME & STEREO-A/COR 2 & First appearance height (time) & 4.84 R$_{\odot}$ (8:39 UT) \\
    CME & STEREO-B/COR 1 & First appearance height (time) & 2.93 R$_{\odot}$ (9:06 UT) \\
    CME & STEREO-B/COR 2 & First appearance height (time) & 3.14 R$_{\odot}$ (8:55 UT) \\ \bottomrule[0.8pt]\midrule[0.3pt]
  \end{tabular} 
  }  

\caption{Timeline of the January 26 2014 CME and the associated phenomena.} \label{table}
\end{sidewaystable}

\section{Data Analysis and Results}\label{results}
\subsection{CME Kinematics from White-Light Data}

\begin{figure}[!ht]  
   \centerline{\hspace*{0.04\textwidth}
               \includegraphics[width=0.6\textwidth,clip=]{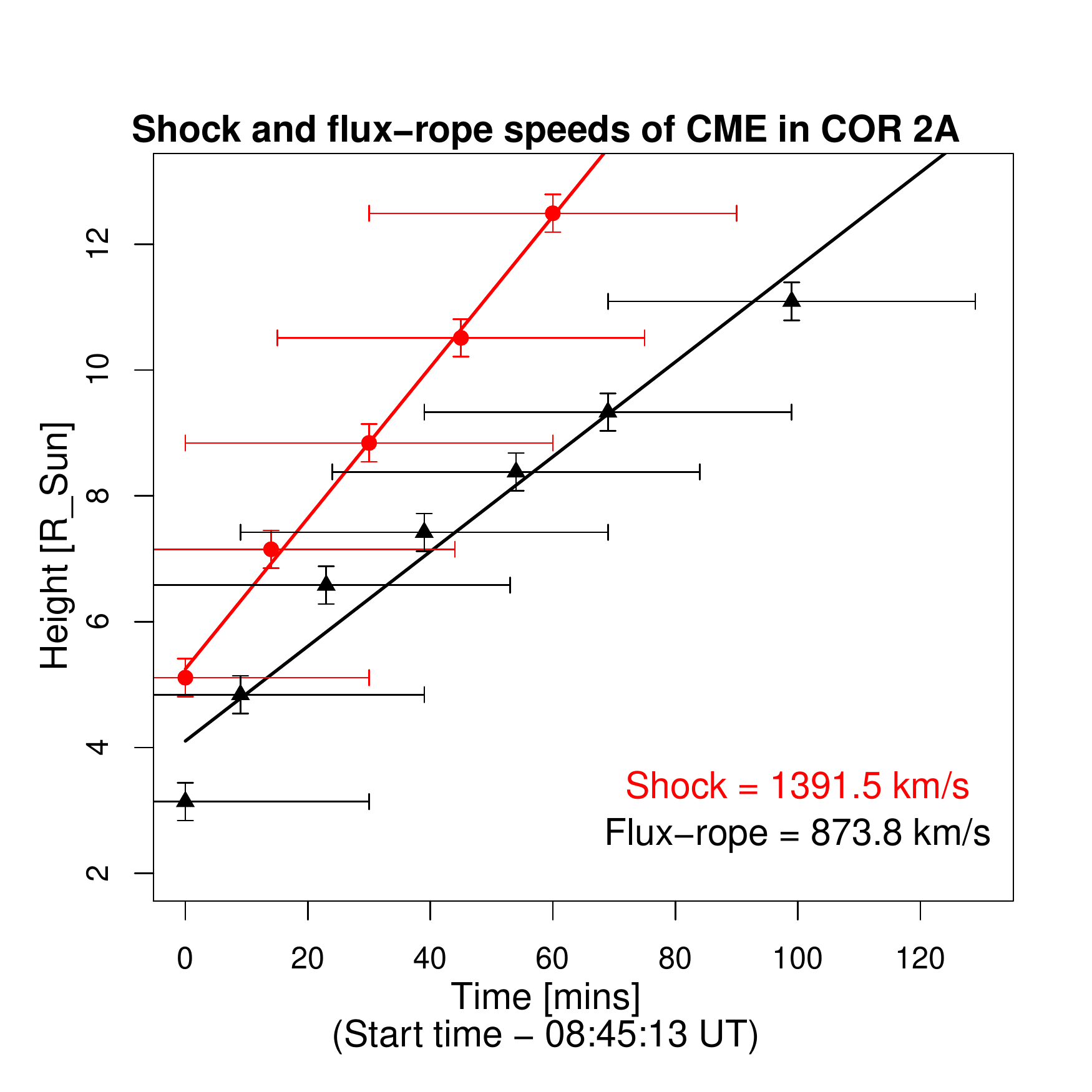}
               \hspace*{-0.02\textwidth}
               \includegraphics[width=0.6\textwidth,clip=]{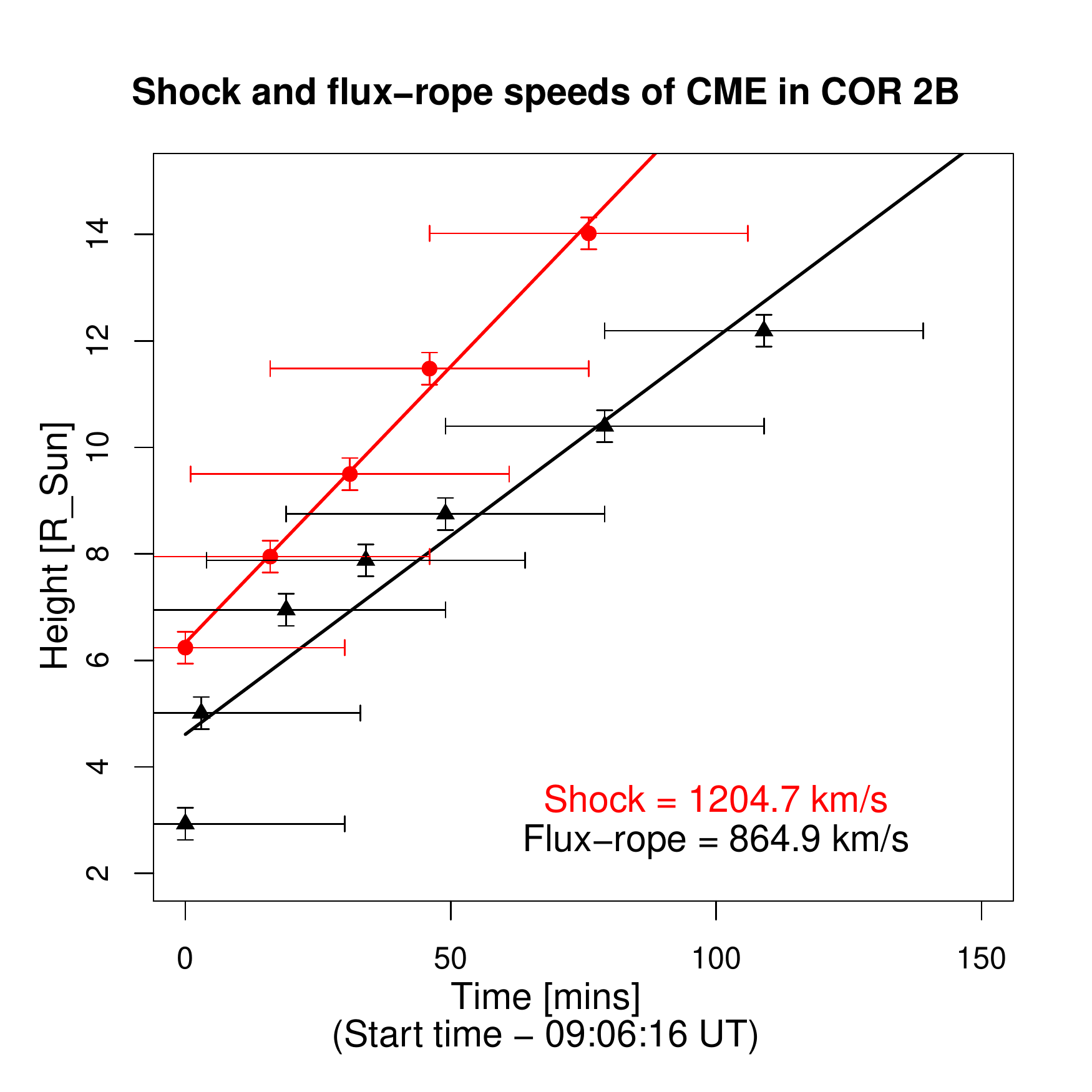}
              }
      \vspace{-0.01\textwidth}  
     \centerline{    
      \hspace{0.2\textwidth}  \color{black}{(a)}
      \hspace{0.58\textwidth}  \color{black}{(b)}
         \hfill}
     \vspace{0.005\textwidth}    
         
 \centerline{\hspace*{0.05\textwidth}
               \includegraphics[width=0.6\textwidth,clip=]{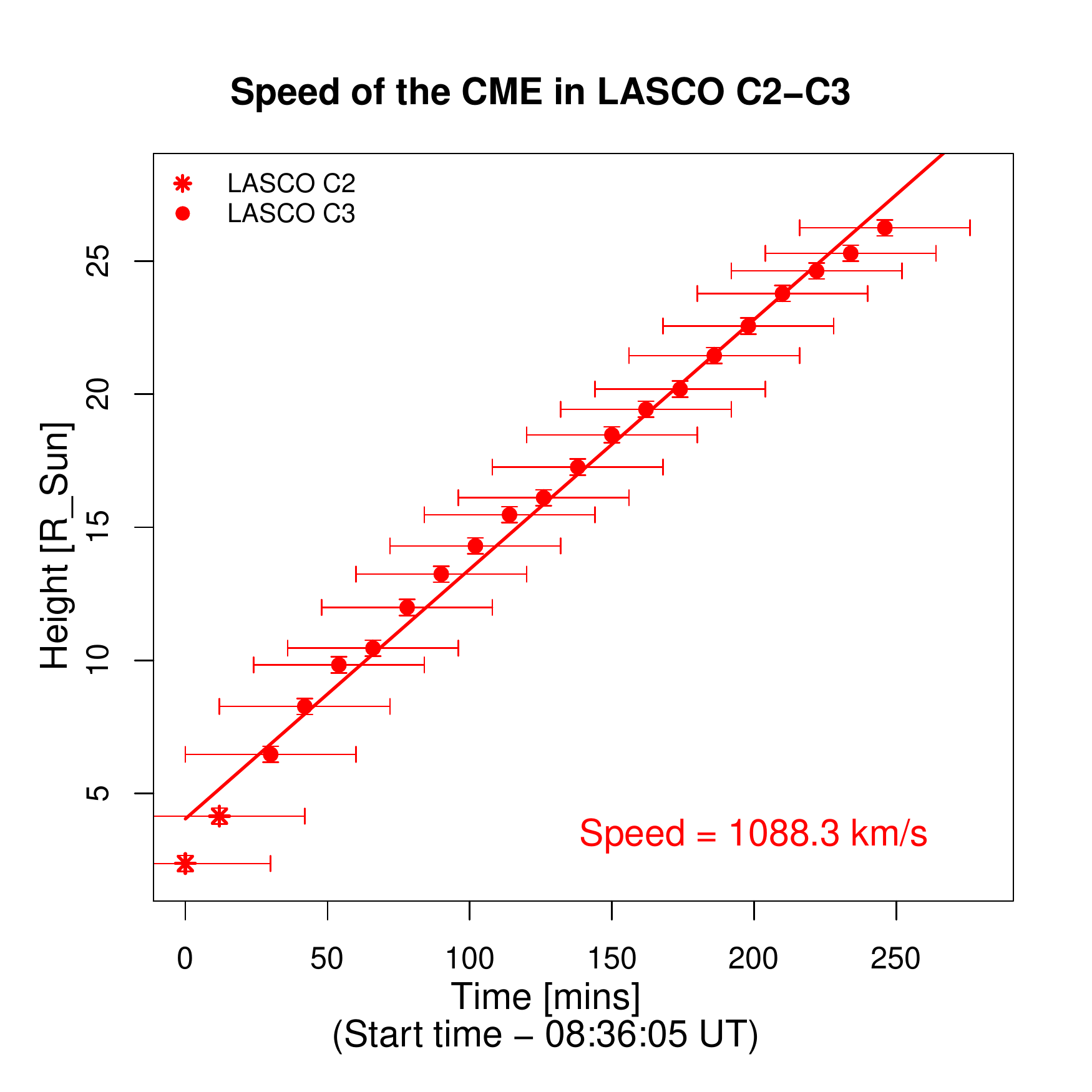}
               \hspace*{0.002\textwidth}
               \includegraphics[width=0.6\textwidth,clip=]{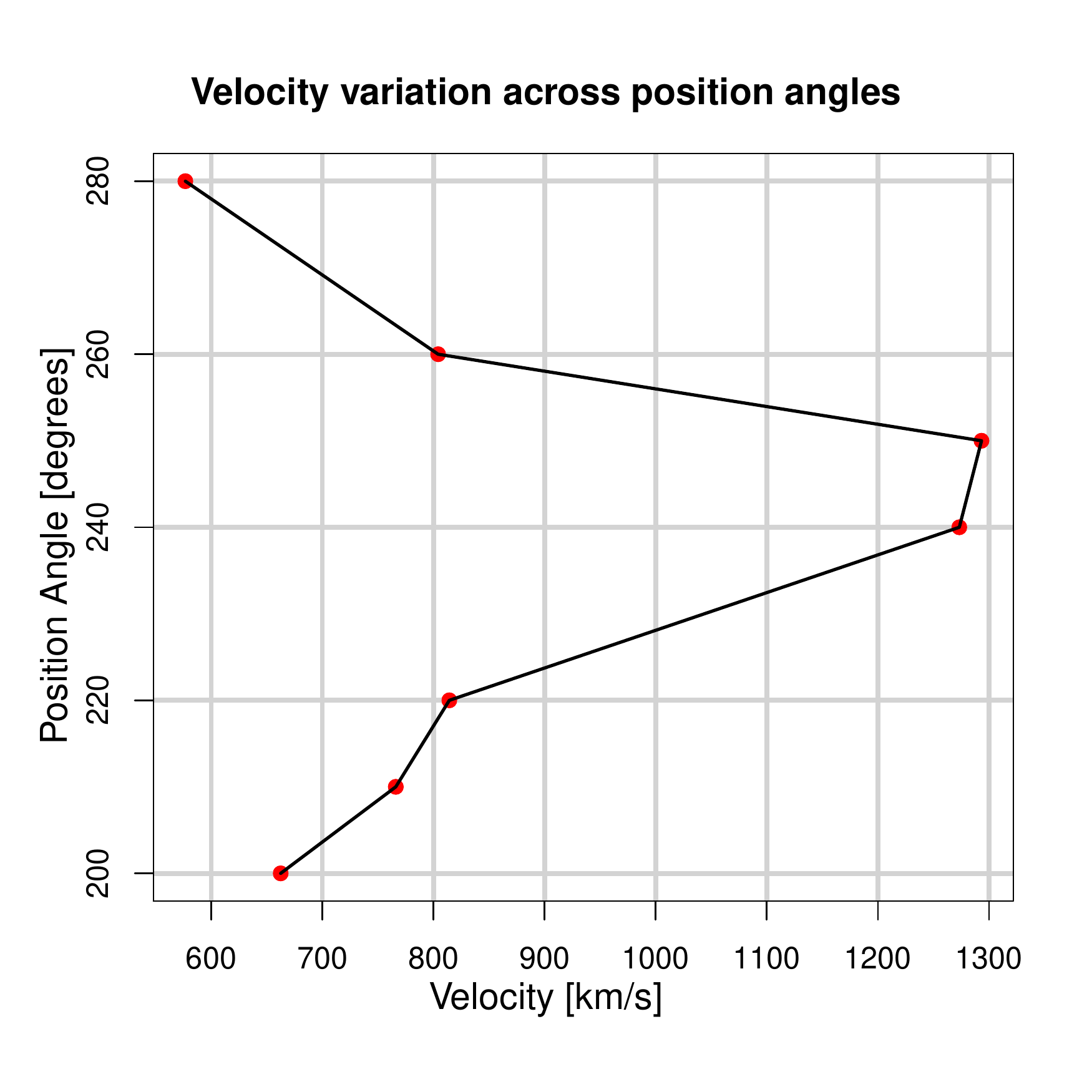}
              }
     \vspace{-0.01\textwidth}  
     \centerline{  
      \hspace{0.2\textwidth}  \color{black}{(c)}
      \hspace{0.58\textwidth}  \color{black}{(d)}
         \hfill}
     \vspace{0.01\textwidth}
         
\caption{The height--time profile of CME shock and flux-rope as observed in (a) STEREO/COR-2A (measured at PA $\approx242^{\circ}$); (b) STEREO/COR-2B (measured at PA $\approx234^{\circ}$). In panel a and b, the red and blue points correspond to the shock front the flux-rope of the CME, respectively. The speed measured at the leading edge of the CME with the SOHO/LASCO data from CDAW catalogue (c). The CME leading edge speed variation with position angles is shown in panel d and the corresponding position angles are marked in Figure~\ref{cme_views}a}.
   \label{fr_shock}
\end{figure}

\begin{figure}[!ht]  
   \centerline{\hspace*{0.04\textwidth}
               \includegraphics[width=0.45\textwidth,clip=]{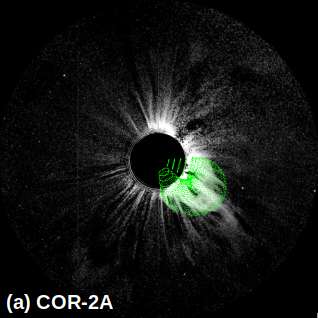}
               \hspace*{0.02\textwidth}
               \includegraphics[width=0.45\textwidth,clip=]{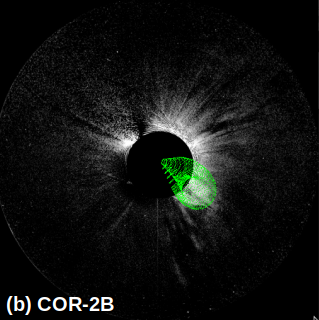}
              }
      \vspace{0.02\textwidth}  
     \centerline{    
      \hspace{0.25\textwidth}  \color{black}{(a)}
      \hspace{0.4\textwidth}  \color{black}{(b)}
         \hfill}
     \vspace{0.005\textwidth}    
         
 \centerline{\hspace*{0.05\textwidth}
              \includegraphics[width=0.55\textwidth,clip=]{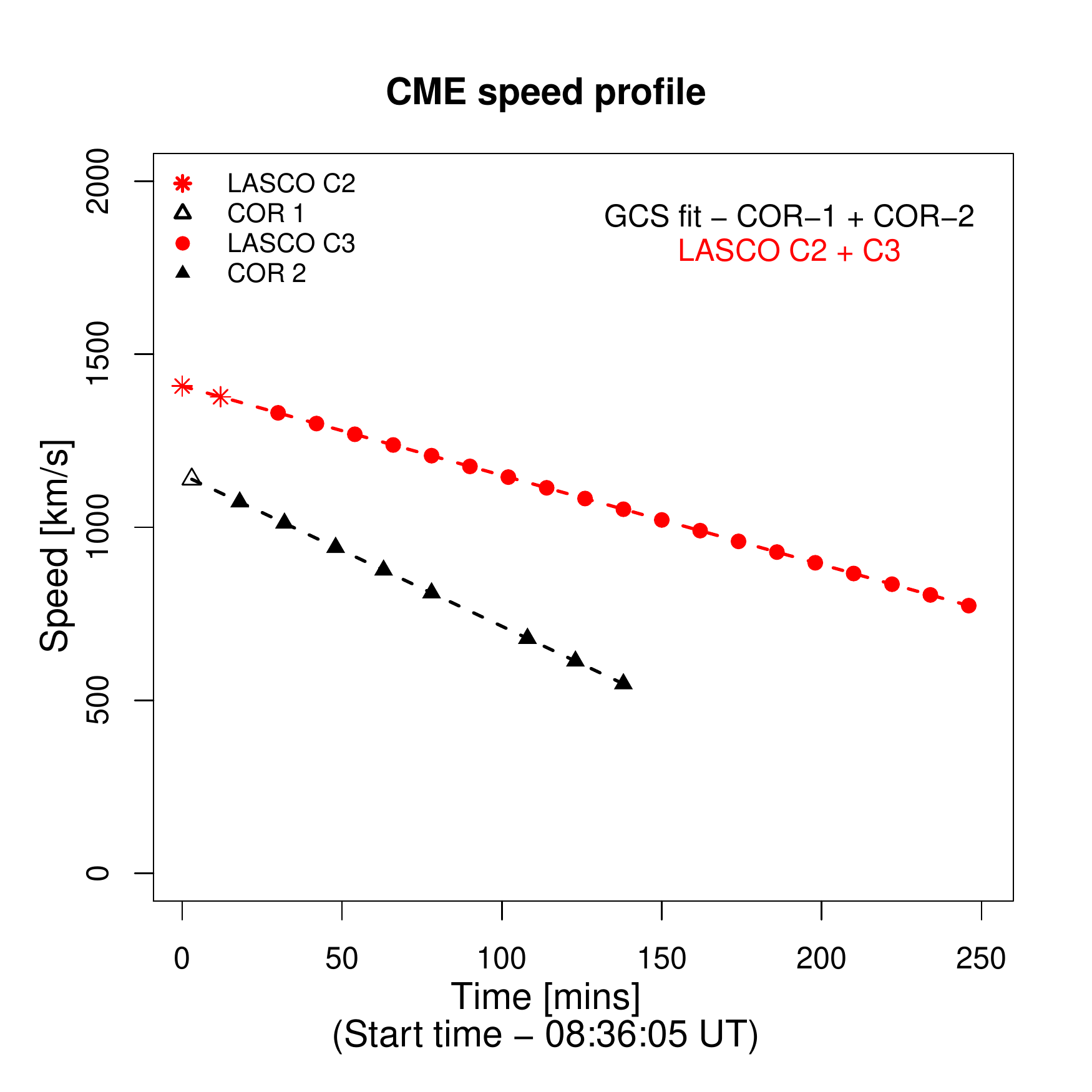}
              }
     \vspace{0.002\textwidth}  
     \centerline{  
      \hspace{0.5\textwidth}  \color{black}{(c)}
      \hfill}
     \vspace{0.01\textwidth}
         
\caption{(a) The CME as observed in COR-2A; (b) COR-2B at 09:24 UT with the GCS model fit in green; (c) The variation of CME leading front speed with time in the LASCO C2 and C3 FOV and from the GCS model fit to the CME in the COR-1 and COR-2 A/B FOV. The CME is clearly decelerating in the coronagraphs FOV.}.
   \label{speed}
\end{figure}

The CME as observed in STEREO/COR-2A, STEREO/COR-2B, and SOHO/ LASCO-C3 is shown in Figure \ref{cme_views}. We track the CME shock front and the flux-rope structure in the COR-2A (measured at PA $\approx242^{\circ}$) and COR-2B (measured at PA $\approx242^{\circ}$) FOV (Figure \ref{fr_shock}a and b, respectively). After repeating the height measurements (for ten times), we found an average measurement error (in height) of 0.3 \,R$_{\odot}$ in COR-2 and LASCO coronagraphs FOV. From linear fits to COR-2A (COR-2B) height--time measurements, we find the average white-light shock speed to be  $\approx$ 1390 \,km s$^{-1}$ (1205 \,km s$^{-1}$) and that of the flux-rope to be 874 \,km s$^{-1}$ (865 \,km s$^{-1}$) (Figure~\ref{fr_shock}a and b). From a linear fit to height--time data from the CDAW catalogue (Figure \ref{fr_shock}c) we find the average speed to be 1088 \,km s$^{-1}$, which is likely to be the white-light shock speed as CDAW tracks the leading edge. Since STEREO-A and LASCO measurements have minimum projection effects (see Section \ref{description}), with correction factors (the ratio of the projected (2D) to the actual (3D) quantities, which varies as $cos(\phi)$ where $\phi$ is the angle from the plane of the sky) of 1.03 and 1.04 respectively, we find the true speeds in COR-2A to be 900 \,km s$^{-1}$ (flux-rope) and 1434 \,km s$^{-1}$ (shock), while the true speed in LASCO is  $\approx$ 1130 \,km s$^{-1}$. For STEREO-B, it is a disk event, and hence the measurements will have large projection effects. With a correction factor  of 1.31, the estimated true speeds in COR-2B are 1133 \,km s$^{-1}$ (flux-rope) and  $\approx$ 1580 \,km s$^{-1}$ (shock). A better way to remove projection effects is to fit the GCS model to the pair of STEREO images to get their true evolution (Figure \ref{speed}a and b). From second-order polynomial fits to height--time measurements in the CDAW catalogue, and from measurements from the GCS model, we find that the CME decelerated in the LASCO and COR-1, COR-2 FOV (Figure \ref{speed}c). We further measured the leading edge of the CME at different Position Angles (PA) in the COR-2A FOV. In Figure \ref{fr_shock}d we plot the average speed of the leading edge versus PA (the corresponding position angles are marked with black dots in Figure~\ref{cme_views}a). We find that the nose of the leading edge travelled with a much higher speed compared to the flanks, and that the southern flank had higher speed than the northern flank. The overall shape of this plot also reproduces the shape of the CME as observed in the coronagraph images (Figure \ref{cme_views}a and c). It is also important to note here that the leading edge of the CME can be the shock front or the flux-rope at different position angles, which may be the reason for the wide range of speeds at different position angles. Thus, we also see that tracking a single point to understand kinematics can be misleading.

\subsection{Connecting Radio and White-Light Data}
\subsubsection{Shock Formation Height} 
  \label{S-text}
It has been reported that Type-II bursts are often associated with CMEs that drive shock waves, and in such cases, the  starting frequency of the Type-II burst can give an estimate of the height of formation of the shock associated with the Type-II burst  \citep{2013AdSpR..51.1981G, kumari2017a} . From the radio dynamic spectrum from the Learmonth station of RSTN, we find the starting frequency of the fundamental Type-II burst is $\approx$ 115 \,MHz at 8:34 UT. \cite{2013AdSpR..51.1981G} found an emperical relation between the shock formation height ($r$) and the starting frequency of the Type-II burst [$f_p$] as follows,  

\begin{equation}
    f = 307.87 r ^{-3.78} - 0.14   \label{eqn1} .
\end{equation} 

\begin{figure}[!ht]  
   \centerline{\hspace*{0.04\textwidth}
               \includegraphics[width=0.6\textwidth,clip=]{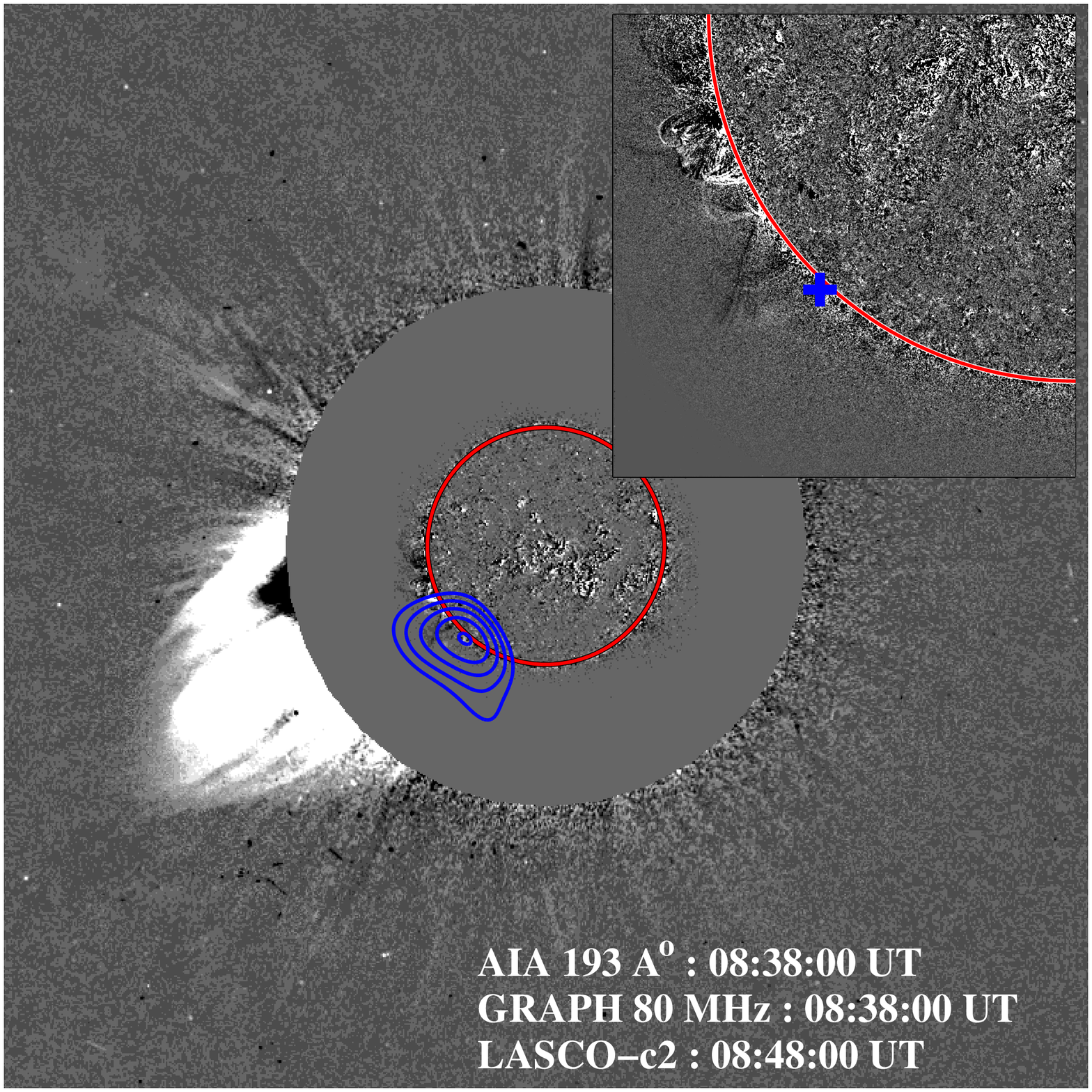}
               }
      \vspace{-0.01\textwidth}  
     \centerline{    
      \hspace{0.05\textwidth}  \color{black}{(a)}
      }
     \vspace{0.005\textwidth}    
         
 \centerline{\hspace*{0.03\textwidth}
               \includegraphics[width=0.5\textwidth,clip=]{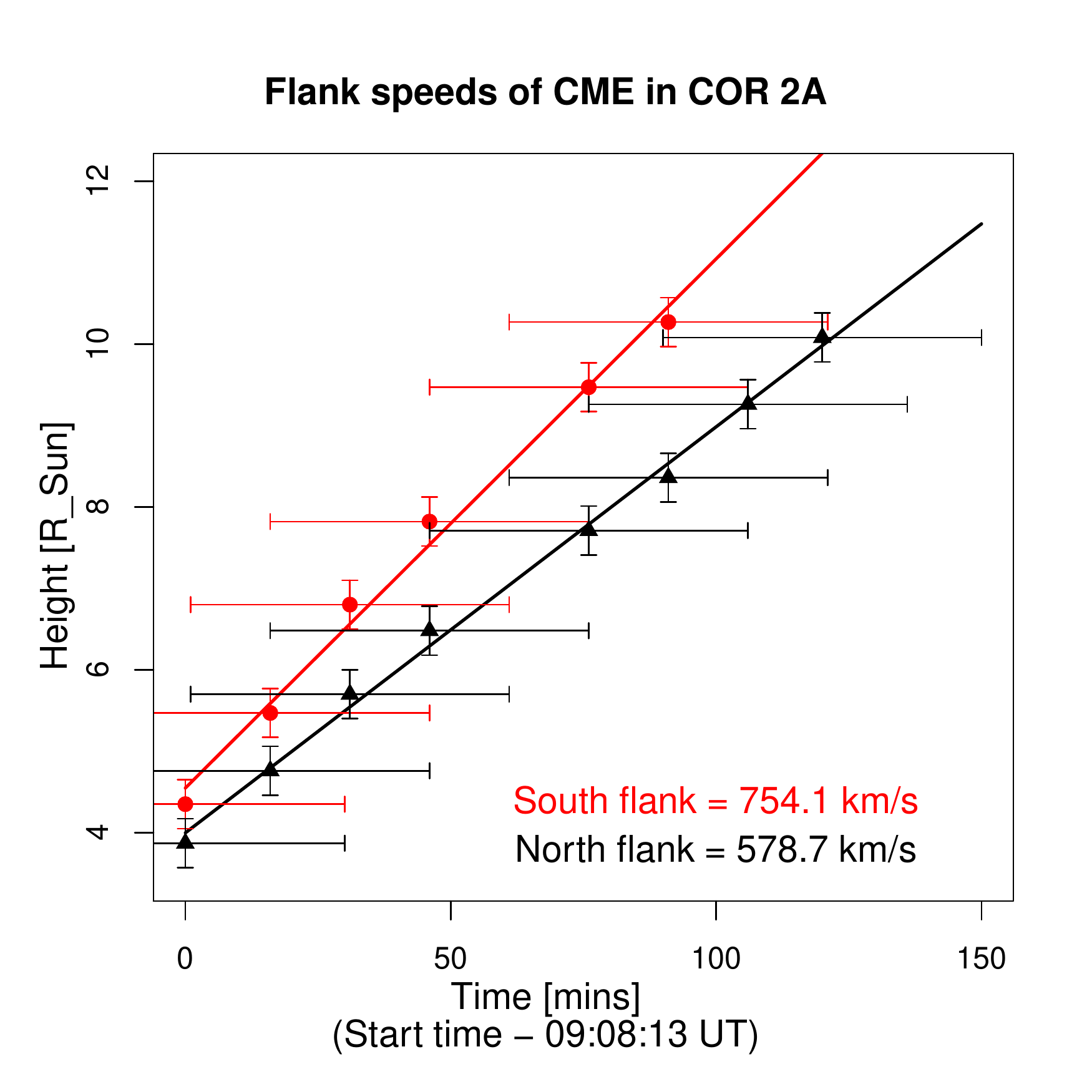}
               \hspace*{0.002\textwidth}
               \includegraphics[width=0.5\textwidth,clip=]{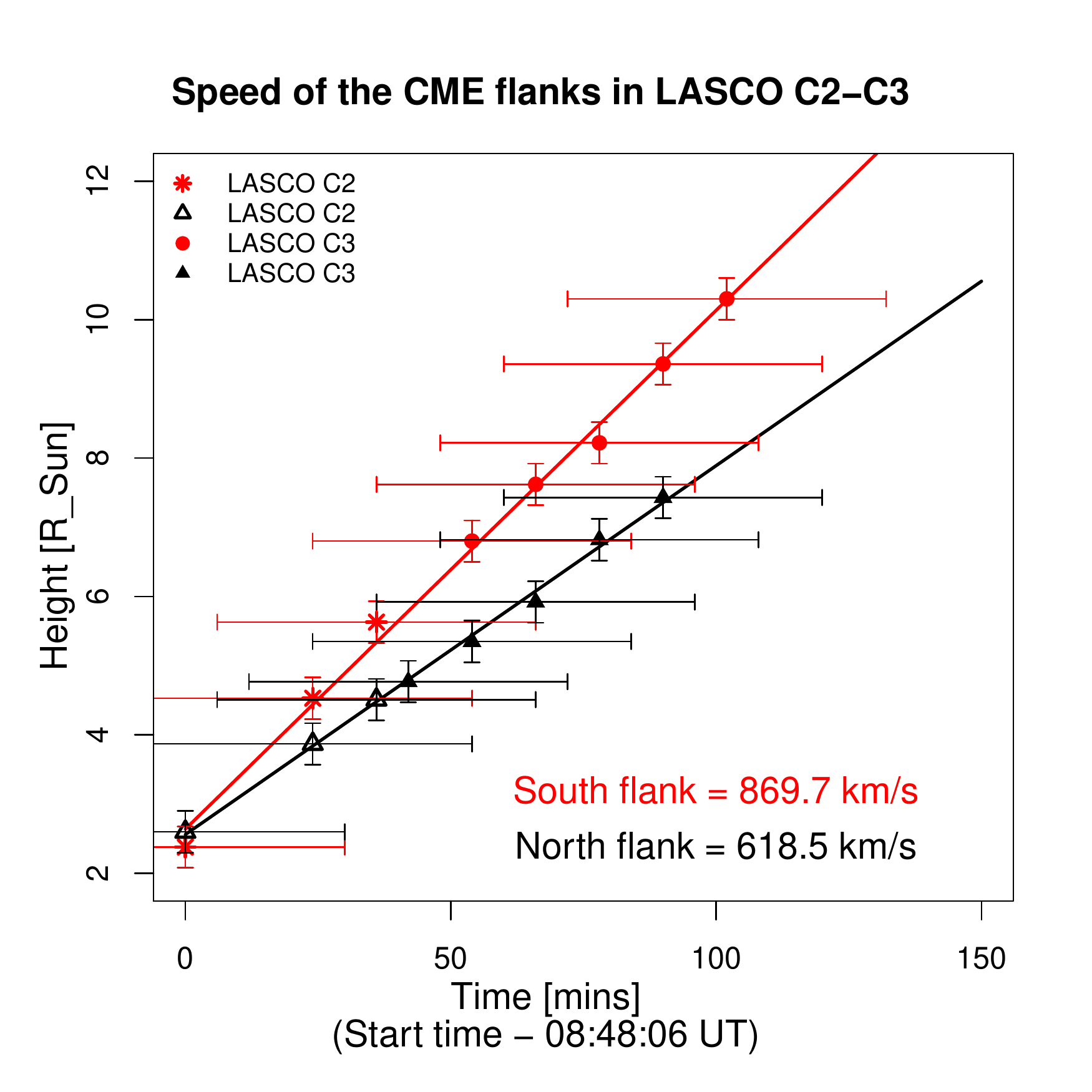}
              }
     \vspace{-0.01\textwidth}  
     \centerline{  
      \hspace{0.24\textwidth}  \color{black}{(b)}
      \hspace{0.45\textwidth}  \color{black}{(c)}
         \hfill}
     \vspace{0.01\textwidth}
         
\caption{(a) The composite of the white-light CME as seen in the SOHO/LASCO-C2 FOV with the 80 MHz contours obtained with GRAPH radio observations and the SDO/AIA-193 {\,\AA}. The red circle marks the solar disk. The radio contours are shown at  50, 72, 75, 87, and 99$\%$ of the peak radio flux at 80 MHz.  The contour intervals are at $\approx 6.6\times10^3$\,Jy. The de-projected height of the radio contours are at $\approx 1.1$R$_{\odot}$ and the position angle is $\approx 157^{\circ}$. An inset with a zoom into the AIA image shows the spatial location of the associated flare (top right).  The location of 80 MHz centroid is marked with blue color in the nset image; The North and South flank speeds of the CME from height--time measurements in (b) COR-2A FOV; and (c) in LASCO C2 and C3 FOV.}
   \label{fig3.2_flank}
\end{figure}

Using the starting frequency in the above relation, we get the height at which the shock wave signatures (Type-II burst) were observed as $\approx 1.3$ R$_{\odot}$ from the Sun center, where R$_{\odot}$ is the radius of the Sun. Since the above relation arises from a weak correlation between the two quantities, we also find the shock-formation height from the relation between the plasma frequency [$f_\mathrm{p}$] and the plasma density [$n_\mathrm{p}$], which is as follows,

\begin{equation}
f_\mathrm{p} = 9 \times10^{-3} \sqrt{n_\mathrm{p}}     \label{eqn2}
\end{equation}

where $f_\mathrm{p}$ is in MHz, The $n_\mathrm{p}$ can be used from the coronal density model. In this work, we use the Newkirk coronal density model \citep{newkirk1967}. We plug in the starting frequency in Equation \ref{eqn2} to get an estimate of the density. With the estimated density, we use the Newkirk coronal density model to 
get the height of Type-II associated shock as $\approx1.3$R$_\odot$, which matches well with what we got from equation \ref{eqn1}. 

From the height--time data obtained with LASCO-C2, the nose of the white-light shock front appears in the FOV at $\approx$ 08:36 UT at a height of $\approx$ $2.33$\,R$_{\odot}$. Taking the instantaneous white-light shock speed from the first two data points in C2 FOV, we calculate the height of the shock front at $\approx$ 8:34 UT as $\approx$ $2.04$ \,R$_{\odot}$. This is higher than the height of the shock associated with the Type-II burst found from the starting frequency of the Type-II burst. In such cases \cite{2009SoPh..259..227G} suggested that the shock resulting in the Type-II burst might come from the electrons accelerated at flanks of the CME (which is at a lower height) and not from the nose. The radio-imaging observation was carried out with GRAPH at 80 MHz (see Figure~\ref{fig3.2_flank}a). The radio image was made at $\approx$  08:38 $\pm$ 20 sec UT, which corresponds to the harmonic band of the Type-II spectra. The radio contours are shown at 50$\,\%$, 72$\,\%$, 75$\,\%$, 87$\,\%$, and 99$\,\%$ of the peak radio flux at 80 \,MHz.  The contour intervals are at $\approx 6.6\times10^3$\,Jy. The projected heliocentric distance for GRAPH is $\approx 1.06$ \,R$_{\odot}$. The active region was located behind the solar disk at $\approx 16^o$ from the east limb. The de-projected heliocentric distance was thus calculated by assuming that the projection effects vary as 1/ cos($\phi$),where $\phi$ is the angle from the plane of sky (POS). The de-projected height of the radio contours was at $\approx 1.1$\,R$_{\odot}$ with a position angle of $\approx 157^{\circ}$, which is at the southern flank of the CME.

We confirm from the GRAPH image (Figure \ref{fig3.2_flank}), that the burst source (produced by the shock wave) located at the southern flank of the CME produces the Type-II burst (please see Section \ref{description} for a description of the imaging data), thus supporting our arguments which are i) The shock formation height inferred from Type-II burst start frequency gives a much lower height than the CME nose height at that time, thus arguing that the shock wave at the flanks of the CME might be the source of the Type-II burst; ii) This is because the flanks are at a lower height than the CME nose, and hence is expected to pass through denser region (as suggested by \cite{2009SoPh..259..227G}). We also note that this was not possible in \cite{2009SoPh..259..227G} to pinpoint which flank of the CME, the emission was coming from, as they did not have imaging observation of the Type-II burst. Additional support to this conclusion comes from the height--time variation of the north and south flanks of the CME in the COR-2A and LASCO-C2,-C3 FOV (Figure~\ref{fig3.2_flank}b and c). The average speeds of the north and south flanks were found to be around 579 \,Km~s$^{-1}$ and 754 \,Km~s$^{-1}$ respectively in the COR-2A FOV, and 619 \,km~s$^{-1}$ and 870 \,km~s$^{-1}$ in the LASCO FOV. This shows that the southern flank is at a higher speed, and thus is likely the source of the Type-II burst.

\subsubsection{Metric Type-II Burst Continuing to DH Domain}    

As mentioned earlier in Section~\ref{description} and from Figure~\ref{typeII} we find that the brief DH Type-II burst is a continuation of the metric Type-II. We note that the DH Type-II burst was observed only in STEREO, not in \textit{Wind}/WAVES, and that it is best observed in STEREO-B because it observes more shock surface. STEREO-A sees only the harmonic. This further supports the flank origin of the Type-II burst. On the other hand, we note that the shock is not fast enough to extend the Type-II emission to the kilometric domain, which happens for shocks at much higher speeds \citep{2005JGRA..11012S07G}. It is interesting that the shock nose, where the speed is the highest, is radio quiet. One possibility is that the local Alf\'{v}en speed above the shock is close to the white-light shock speed, rendering the Mach number to be too small (see, e.g., \cite{2008ApJ...674..560G}). 


\subsubsection{Alf\'{v}en Speed from Stand-off Distance of CME Driven Shock and Band-Split Measurement}

Despite the ending of the Type-II burst at $\approx$ 08:52 UT, we found the shock to be present in COR-2A images as the CME propagated further into the heliosphere. We follow \cite{2011ApJ...736L..17G} to get the Alf\'{v}en speed at the nose around the time the Type-II burst ended. In the 09:24 UT COR-2A image, we measure the standoff distance [$\Delta R$] as $\approx$ 0.99 \,R$_{\odot}$ and a radius of curvature ($R_\mathrm{c}$) of $\approx$ 1.34 \,R$_{\odot}$ (at a height of $\approx$ 6 \,R$_{\odot}$). Using these values in the \cite{RUSSELL2002527} relation,

\begin{equation}
    \frac{\Delta R}{R_\mathrm{c}} = 0.81\frac{(\gamma - 1)M^2 + 2}{(\gamma + 1)(M^2 - 1)},
\end{equation}
and the adiabatic index $\gamma$ as $5/3$ we get the Mach number [$M$] as $\approx$ 1.31. Using the local white-light shock speed as 1250 \,km s$^{-1}$ in the relation,

\begin{equation}
    V_\mathrm{A} = \frac{V_{\mathrm{shock}}}{M}, \label{eqn5}
\end{equation}

we get an Alf\'{v}en speed of $\approx$ 954 km~s$^{-1}$. This is higher than the typical Alf\'{v}en speed at these heights \citep{vrsnak_2004}, thus confirming weak shock conditions at the nose of the CME. We also use the band splitting of the Type-II burst to get the Alf\'{v}en speed. We follow a similar procedure to that reported by \citet{vrsnak_2001}. We measure the width of the band splitting of the fundamental Type-II burst on the dynamic spectrum (Figure~\ref{typeII}).The bandwidth is defined as

\begin{equation}
    BW = (F_\mathrm{u} -F_\mathrm{l})/F_\mathrm{l},
\end{equation}

where $F_\mathrm{u}$ and $F_\mathrm{l}$ are the upper- and lower-band frequencies respectively. The density jump [$X$] across the shock front is related to the bandwidth as

\begin{equation}
    X = (BW + 1)^2.
\end{equation}

This density jump [$X$] is further related to the Alf\'{v}enic Mach number [$M_\mathrm{A}$] through

\begin{equation}
    M_\mathrm{A} = \sqrt{X(X+5)/2(4-X)}.
\end{equation}
Using the above relation, we get an average $M_\mathrm{A}$ of 1.45. From the heights inferred from the lower band of the fundamental branch of the Type-II burst, we get an average shock speed associated with the Type-II burst as 1451 \,km s$^{-1}$, which from Equation \ref{eqn5} gives us an Alf\'{v}en speed of 1001 \,km s$^{-1}$. It is interesting that although the Alf\'{v}en speed is comparable at the CME nose at 6 \,R$_{\odot}$ and from the band-split calculations, yet the Type-II burst is coming from the flank of the CME (Figure~\ref{fig3.2_flank}a).

\subsubsection{The CME--Streamer Encounter}

\begin{figure}[h]
    \centering
    \includegraphics[scale=0.4]{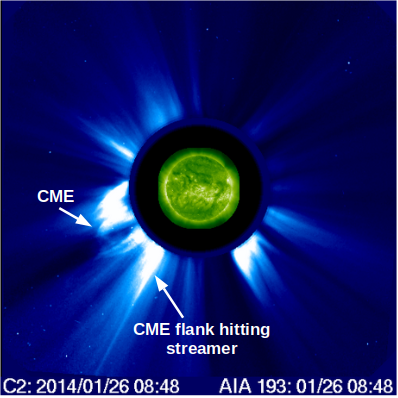}
    \caption{CME interacting with the streamer at the southern flank as depicted in the LASCO-C2 image at 08:48 UT. The LASCO image is superposed on with SDO/AIA 193 \,\AA~ image.}
    \label{fig3.2_cmestream}
\end{figure}

The sudden broadening in the DH Type-II burst seems to be due to the CME interaction with the streamer (as also reported earlier by \cite{feng2012}) at the southern flank (see left panel of Figure~\ref{typeII}). Movies of LASCO-C2 images show the interaction of the CME with the southern streamer at the time when the Type-II broadening was observed around 8:48 UT as shown in Figure \ref{fig3.2_cmestream}, just before the end of the Type-II. The region of interaction was also determined from LASCO movies, where it was seen that the left flank of the CME interacts with the streamer during its propagation. Since the spectral bump in the SWAVES dynamic spectra (Figure~\ref{typeII}) occurred around the same time, the CME flank interacted with the streamer ($\approx$ 8:48 UT), we estimated the height at which the interaction takes place from LASCO images to be $\approx$ 2.58 \,R$_{\odot}$, not too different from the expected height ($\approx$ 2.61 \,R$_{\odot}$) from Equation (\ref{eqn1}). When the DH Type-II ended around 08:42 UT, the shock seems to have transited through the streamer. At 08:52 UT, when the Type-II ended, the nose was at a height of 5 \,R$_{\odot}$, where the shock was not strong enough to produce the Type-II burst. 

Type-II imaging observation confirmed that the source of the Type-II burst is located near the southern flank of the CME (Figure~\ref{fig3.2_flank}). Since the DH Type-II burst is a continuation of the metric Type-II, the emission continues at the southern flank as the shock moves out. The flank speeds are of $\approx$ 574 \,km~s$^{-1}$ (north flank) and $\approx$ 754 \,km~s$^{-1}$ (south flank) as derived by a linear fit to the distance-time plot of flanks shown in Figure \ref{fig3.2_flank}. The speeds are not too different, yet only the southern flank has the source.  At this time, the shock nose is at a higher speed on one hand, and on the other hand the presence of a streamer near the southern flank seems to have provided additional favorable condition for the generation of shock-associated radio emission, because the streamers are denser than the ambient corona and are known to be regions of low Alf\'{v}en speed compared to the normal corona. Also, the nose is at a height of $\approx$2 \,R$_{\odot}$, where the Alf\'{v}en speed is near its peak value \citep{vrsnak_2004}, so the shock is relatively weaker at the nose to produce the Type-II burst. Thus the combination of white-light and radio imaging along with radio dynamic spectrum provides conclusive evidence that the CME-streamer interaction is also additionally responsible for the generation of shock associated Type-II emission.

\section{Summary and Conclusions}\label{summary}

The primary finding of this article is that we were able to show that the Type-II burst during the 26 January 2014 CME originated from one of the flanks of the CME-driven shock. We combined EUV, radio, and white-light data from various space- and ground-based instruments to understand the kinematic aspects associated with the CME including true speed, shock propagation speed, Alfv\'en speed and the association with Type-II bursts, that confirm this conclusion. We were also able to show that the nose region of the CME was radio quiet because of the high Alf\'{v}en speed, hence resulting in a Mach number very close to one there. We summarise in the following points, our main results from this work that supports our conclusion,

\begin{itemize}
    
\item We measured the average speeds of the flux-rope and the shock front and found a substantial difference in their speeds. The stand-off distance of the shock increases with height, consistent with the deceleration of the CME in the coronagraph FOV. 
    
\item The CME speeds measured at different position angles reveal that the nose was the fastest. However, the flanks are also fast enough to drive a shock and produce Type-II radio emission. Even though the northern and southern flanks had similar speeds, the southern flank interacted with a streamer, which may be the reason that the Type-II burst originated from the southern flank. From LASCO-C2 data, the shock height was estimated to be at $\approx$~2.04\,R$_{\odot}$, which was much higher than the shock-formation height inferred from the starting frequency of Type-II burst ($\approx$ $1.3$ \,R$_{\odot}$), thus suggesting that the Type-II burst signal might be coming from the CME flanks. A plot of the average flank speeds showed that the southern flank was at a higher speed than the northern flank. Further, from the radio contours from GRAPH it was evident that the burst source was coming from the southern flank of the CME. It is also worthwhile to note that the radio imaging observation of the Type-II burst enabled us in pinpointing the flank emission of the Type-II burst, which was not possible in \cite{2009SoPh..259..227G}.

\item It was interesting that the shock nose, where the speed was highest, did not produce the Type-II emission. From the standoff-distance measurements at the CME nose, we found a higher Alf\'{v}en speed, confirming a weak shock there. Also, the presence and interaction of the streamer with the southern flank of the CME seems to have provided more favorable conditions for a strong shock at the flanks, thus validating our conclusions.

\end{itemize}

This work shows the importance of complementing spectral radio data with imaging data in locating the part of the CME driven shock responsible for accelerating electrons. An understanding of the location and origin of the shock waves that are associated with Type-II emission is a complex problem especially in the inner corona (see \citet{2020A&A...639A..56J} and references therein). Further, since shocks accelerate particles that affects space weather, it becomes necessary to study and identify their origin, and this work particularly aims at improving our present understanding of the same. It should also be noted here that we obtained the shock formation height of $\approx$1.3 \,R$_\odot$ using the radio data. This height is below the existing space-based white-light coronagraphs used for analysis. It is worth mentioning that future space-based missions including \textit{Aditya-L1} \citep{ADITYA2017, VELC17}, PROBA-3 \citep{Proba3}, and \textit{Solar Orbiter} \citep{solarorbiter}, are equipped with coronagraphs capable of observing the inner corona, a region with limited observations from existing space-based instruments. These instruments will be helpful in studies similar to this work to identify the white-light counterparts of the radio imaging of the shock origin, thereby improving understanding of such phenomenon.


\begin{acks}
We thank the anonymous reviewer for the valuable comments that have improved the manuscript. This work was done during the hands on data sessions of the COSPAR Capacity Building Workshop on Coronal and Interplanetary Shocks: Analysis of Data from Space and Ground Based Instruments held at Kodaikanal Solar Observatory, Tamil Nadu, India during 6\,--\,17 January 2020. The authors thank the organisers of the workshop for providing the opportunity to learn and work with experts in the field. We also extend our gratitude to Christian Monstein, Seiji Yashiro and Indrajit V. Barve for developing the Python codes, which helped us a lot during the data analysis part of the workshop. We also take this opportunity to thank the local organisers who took care of the local logistics, which enabled a smooth running of the workshop. We  would  like  to  express  our gratitude to the Gauribidanur Radio Observatory staff for providing the radio heliograph data. The SOHO/LASCO data used here are produced by a consortium of the Naval Research Laboratory (USA), Max-Planck-Institut f\"ur Aeronomie (Germany), Laboratoire d'Astronomie (France), and the University of Birmingham (UK). SOHO is a project of international cooperation between ESA and NASA. The SECCHI data used here were produced by an international consortium of the Naval Research Laboratory (USA), Lockheed Martin Solar and Astrophysics Lab (USA), NASA Goddard Space Flight Center (USA), Rutherford Appleton Laboratory (UK), University of Birmingham (UK), Max-Planck-Institut for Solar System Research (Germany), Centre Spatiale de Liège (Belgium), Institut d’Optique Théorique et Appliquée (France), Institut d’Astrophysique Spatiale (France). We also acknowledge SDO team for making the AIA data available. SDO is a mission for NASA’s Living With a Star (LWS) program. S.Majumdar acknowledges Dipankar Banerjee for his constant support and motivation to participate in this workshop. A.Kumari acknowledges the ERC under the European Union's Horizon 2020 Research and Innovation Programme Project SolMAG 724391. N.Gopalswamy was supported by NASA's Living With a Star program. 
\end{acks}

{\footnotesize
\paragraph*{Disclosure of Potential Conflicts of Interest}
The authors declare that they have no conflict of interest.
}




\bibliographystyle{spr-mp-sola1}

\begin{thebibliography}{59}
\ifx\bisbn     \undefined \def\bisbn  #1{ISBN #1}\fi
\ifx\binits    \undefined \def\binits#1{#1}\fi
\ifx\bauthor   \undefined \def\bauthor#1{#1}\fi
\ifx\batitle   \undefined \def\batitle#1{#1}\fi
\ifx\bjtitle   \undefined \def\bjtitle#1{\textit{#1}}\fi
\ifx\bvolume   \undefined \def\bvolume#1{\textbf{#1}}\fi
\ifx\byear     \undefined \def\byear#1{#1}\fi
\ifx\bissue    \undefined \def\bissue#1{#1}\fi
\ifx\bfpage    \undefined \def\bfpage#1{#1}\fi
\ifx\blpage    \undefined \def\blpage #1{#1}\fi
\ifx\burl      \undefined \def\burl#1{\textsf{#1}}\fi
\ifx\href      \undefined \def\href#1#2{\textsf{#2}}\fi
\ifx\betal     \undefined \def\betal{\textit{et al.}}\fi
\ifx\bctitle   \undefined \def\bctitle#1{#1}\fi
\ifx\beditor   \undefined \def\beditor#1{#1}\fi
\ifx\bbtitle   \undefined \def\bbtitle#1{\textit{#1}}\fi
\ifx\bedition  \undefined \def\bedition#1{#1}\fi
\ifx\bseriesno \undefined \def\bseriesno#1{\textbf{#1}}\fi
\ifx\blocation \undefined \def\blocation#1{#1}\fi
\ifx\bsertitle \undefined \def\bsertitle#1{\textit{#1}}\fi
\ifx\bsnm      \undefined \def\bsnm#1{#1}\fi
\ifx\bsuffix   \undefined \def\bsuffix#1{#1}\fi
\ifx\bparticle \undefined \def\bparticle#1{#1}\fi
\ifx\barticle  \undefined \def\barticle#1{}\fi
\ifx\binstitute  \undefined \def\binstitute#1{#1}\fi
\ifx\bpublisher  \undefined \def\bpublisher#1{#1}\fi
\ifx\doiurl    \undefined
  \def\doiurl#1{\href{http://dx.doi.org/#1}{\textsf{DOI}}}\fi
\ifx\arxivurl  \undefined
  \def\arxivurl#1{\href{http://arxiv.org/abs/#1}{\textsf{arXiv}}}\fi
\ifx\adsurl    \undefined
  \def\adsurl#1{\href{http://adsabs.harvard.edu/abs/#1}{\textsf{ADS}}}\fi
\ifx\botherref \undefined \def\botherref#1{}\fi
\ifx\url       \undefined \def\url#1{\textsf{#1}}\fi
\ifx\bchapter  \undefined \def\bchapter#1{}\fi
\ifx\bbook     \undefined \def\bbook#1{}\fi
\ifx\bcomment  \undefined \def\bcomment#1{#1}\fi
\ifx\oauthor   \undefined \def\oauthor#1{#1}\fi
\ifx\citeauthoryear \undefined\def \citeauthoryear#1{#1}\fi
\ifx\endbibitem\undefined \def\endbibitem{}\fi
\ifx\bconflocation  \undefined \def\bconflocation#1{#1} \fi

\bibitem[\protect\citeauthoryear{{Bale} et~al.}{1999}]{1999GeoRL..26.1573B}
\begin{barticle}
\bauthor{\bsnm{{Bale}}, \binits{S.D.}},
\bauthor{\bsnm{{Reiner}}, \binits{M.J.}},
\bauthor{\bsnm{{Bougeret}}, \binits{J.-L.}},
\bauthor{\bsnm{{Kaiser}}, \binits{M.L.}},
\bauthor{\bsnm{{Krucker}}, \binits{S.}},
\bauthor{\bsnm{{Larson}}, \binits{D.E.}},
\bauthor{\bsnm{{Lin}}, \binits{R.P.}}:
\byear{1999},
\batitle{{The source region of an interplanetary type II radio burst}}.
\bjtitle{\grl}
\bvolume{26}(\bissue{11}),
\bfpage{1573}.
\doiurl{10.1029/1999GL900293}.
\adsurl{https://ui.adsabs.harvard.edu/abs/1999GeoRL..26.1573B}.
\end{barticle}
\endbibitem

\bibitem[\protect\citeauthoryear{{Balmaceda}
  et~al.}{2018}]{2018ApJ...863...57B}
\begin{barticle}
\bauthor{\bsnm{{Balmaceda}}, \binits{L.A.}},
\bauthor{\bsnm{{Vourlidas}}, \binits{A.}},
\bauthor{\bsnm{{Stenborg}}, \binits{G.}},
\bauthor{\bsnm{{Dal Lago}}, \binits{A.}}:
\byear{2018},
\batitle{{How Reliable Are the Properties of Coronal Mass Ejections Measured
  from a Single Viewpoint?}}
\bjtitle{\apj}
\bvolume{863}(\bissue{1}),
\bfpage{57}.
\doiurl{10.3847/1538-4357/aacff8}.
\adsurl{https://ui.adsabs.harvard.edu/abs/2018ApJ...863...57B}.
\end{barticle}
\endbibitem

\bibitem[\protect\citeauthoryear{{Bougeret} et~al.}{2008}]{SWAVES}
\begin{barticle}
\bauthor{\bsnm{{Bougeret}}, \binits{J.L.}},
\bauthor{\bsnm{{Goetz}}, \binits{K.}},
\bauthor{\bsnm{{Kaiser}}, \binits{M.L.}},
\bauthor{\bsnm{{Bale}}, \binits{S.D.}},
\bauthor{\bsnm{{Kellogg}}, \binits{P.J.}},
\bauthor{\bsnm{{Maksimovic}}, \binits{M.}},
\bauthor{\bsnm{{Monge}}, \binits{N.}},
\bauthor{\bsnm{{Monson}}, \binits{S.J.}},
\bauthor{\bsnm{{Astier}}, \binits{P.L.}},
\bauthor{\bsnm{{Davy}}, \binits{S.}},
\bauthor{\bsnm{{Dekkali}}, \binits{M.}},
\bauthor{\bsnm{{Hinze}}, \binits{J.J.}},
\bauthor{\bsnm{{Manning}}, \binits{R.E.}},
\bauthor{\bsnm{{Aguilar-Rodriguez}}, \binits{E.}},
\bauthor{\bsnm{{Bonnin}}, \binits{X.}},
\bauthor{\bsnm{{Briand}}, \binits{C.}},
\bauthor{\bsnm{{Cairns}}, \binits{I.H.}},
\bauthor{\bsnm{{Cattell}}, \binits{C.A.}},
\bauthor{\bsnm{{Cecconi}}, \binits{B.}},
\bauthor{\bsnm{{Eastwood}}, \binits{J.}},
\bauthor{\bsnm{{Ergun}}, \binits{R.E.}},
\bauthor{\bsnm{{Fainberg}}, \binits{J.}},
\bauthor{\bsnm{{Hoang}}, \binits{S.}},
\bauthor{\bsnm{{Huttunen}}, \binits{K.E.J.}},
\bauthor{\bsnm{{Krucker}}, \binits{S.}},
\bauthor{\bsnm{{Lecacheux}}, \binits{A.}},
\bauthor{\bsnm{{MacDowall}}, \binits{R.J.}},
\bauthor{\bsnm{{Macher}}, \binits{W.}},
\bauthor{\bsnm{{Mangeney}}, \binits{A.}},
\bauthor{\bsnm{{Meetre}}, \binits{C.A.}},
\bauthor{\bsnm{{Moussas}}, \binits{X.}},
\bauthor{\bsnm{{Nguyen}}, \binits{Q.N.}},
\bauthor{\bsnm{{Oswald}}, \binits{T.H.}},
\bauthor{\bsnm{{Pulupa}}, \binits{M.}},
\bauthor{\bsnm{{Reiner}}, \binits{M.J.}},
\bauthor{\bsnm{{Robinson}}, \binits{P.A.}},
\bauthor{\bsnm{{Rucker}}, \binits{H.}},
\bauthor{\bsnm{{Salem}}, \binits{C.}},
\bauthor{\bsnm{{Santolik}}, \binits{O.}},
\bauthor{\bsnm{{Silvis}}, \binits{J.M.}},
\bauthor{\bsnm{{Ullrich}}, \binits{R.}},
\bauthor{\bsnm{{Zarka}}, \binits{P.}},
\bauthor{\bsnm{{Zouganelis}}, \binits{I.}}:
\byear{2008},
\batitle{{S/WAVES: The Radio and Plasma Wave Investigation on the STEREO
  Mission}}.
\bjtitle{\ssr}
\bvolume{136}(\bissue{1-4}),
\bfpage{487}.
\doiurl{10.1007/s11214-007-9298-8}.
\adsurl{https://ui.adsabs.harvard.edu/abs/2008SSRv..136..487B}.
\end{barticle}
\endbibitem

\bibitem[\protect\citeauthoryear{{Brueckner} et~al.}{1995}]{Brueckner95}
\begin{barticle}
\bauthor{\bsnm{{Brueckner}}, \binits{G.E.}},
\bauthor{\bsnm{{Howard}}, \binits{R.A.}},
\bauthor{\bsnm{{Koomen}}, \binits{M.J.}},
\bauthor{\bsnm{{Korendyke}}, \binits{C.M.}},
\bauthor{\bsnm{{Michels}}, \binits{D.J.}},
\bauthor{\bsnm{{Moses}}, \binits{J.D.}},
\bauthor{\bsnm{{Socker}}, \binits{D.G.}},
\bauthor{\bsnm{{Dere}}, \binits{K.P.}},
\bauthor{\bsnm{{Lamy}}, \binits{P.L.}},
\bauthor{\bsnm{{Llebaria}}, \binits{A.}},
\bauthor{\bsnm{{Bout}}, \binits{M.V.}},
\bauthor{\bsnm{{Schwenn}}, \binits{R.}},
\bauthor{\bsnm{{Simnett}}, \binits{G.M.}},
\bauthor{\bsnm{{Bedford}}, \binits{D.K.}},
\bauthor{\bsnm{{Eyles}}, \binits{C.J.}}:
\byear{1995},
\batitle{{The Large Angle Spectroscopic Coronagraph (LASCO)}}.
\bjtitle{\solphys}
\bvolume{162},
\bfpage{357}.
\doiurl{10.1007/BF00733434}.
\adsurl{1995SoPh..162..357B}.
\end{barticle}
\endbibitem

\bibitem[\protect\citeauthoryear{{Cane} and {Stone}}{1984}]{cane1984type}
\begin{barticle}
\bauthor{\bsnm{{Cane}}, \binits{H.V.}},
\bauthor{\bsnm{{Stone}}, \binits{R.G.}}:
\byear{1984},
\batitle{{Type II solar radio bursts, interplanetary shocks, and energetic
  particle events}}.
\bjtitle{Astrophys. J.}
\bvolume{282},
\bfpage{339}.
\doiurl{10.1086/162207}.
\adsurl{https://ui.adsabs.harvard.edu/abs/1984ApJ...282..339C}.
\end{barticle}
\endbibitem

\bibitem[\protect\citeauthoryear{{Cho} et~al.}{2007}]{2007ApJ...665..799C}
\begin{barticle}
\bauthor{\bsnm{{Cho}}, \binits{K.-S.}},
\bauthor{\bsnm{{Lee}}, \binits{J.}},
\bauthor{\bsnm{{Gary}}, \binits{D.E.}},
\bauthor{\bsnm{{Moon}}, \binits{Y.-J.}},
\bauthor{\bsnm{{Park}}, \binits{Y.D.}}:
\byear{2007},
\batitle{{Magnetic Field Strength in the Solar Corona from Type II Band
  Splitting}}.
\bjtitle{\apj}
\bvolume{665}(\bissue{1}),
\bfpage{799}.
\doiurl{10.1086/519160}.
\adsurl{https://ui.adsabs.harvard.edu/abs/2007ApJ...665..799C}.
\end{barticle}
\endbibitem

\bibitem[\protect\citeauthoryear{{Cho} et~al.}{2013}]{cho2013high}
\begin{barticle}
\bauthor{\bsnm{{Cho}}, \binits{K.-S.}},
\bauthor{\bsnm{{Gopalswamy}}, \binits{N.}},
\bauthor{\bsnm{{Kwon}}, \binits{R.-Y.}},
\bauthor{\bsnm{{Kim}}, \binits{R.-S.}},
\bauthor{\bsnm{{Yashiro}}, \binits{S.}}:
\byear{2013},
\batitle{{A High-frequency Type II Solar Radio Burst Associated with the 2011
  February 13 Coronal Mass Ejection}}.
\bjtitle{Astrophys. J.}
\bvolume{765}(\bissue{2}),
\bfpage{148}.
\doiurl{10.1088/0004-637X/765/2/148}.
\adsurl{https://ui.adsabs.harvard.edu/abs/2013ApJ...765..148C}.
\end{barticle}
\endbibitem

\bibitem[\protect\citeauthoryear{{Delaboudini{\`e}re} et~al.}{1995}]{SOHOEIT}
\begin{barticle}
\bauthor{\bsnm{{Delaboudini{\`e}re}}, \binits{J.-P.}},
\bauthor{\bsnm{{Artzner}}, \binits{G.E.}},
\bauthor{\bsnm{{Brunaud}}, \binits{J.}},
\bauthor{\bsnm{{Gabriel}}, \binits{A.H.}},
\bauthor{\bsnm{{Hochedez}}, \binits{J.F.}},
\bauthor{\bsnm{{Millier}}, \binits{F.}},
\bauthor{\bsnm{{Song}}, \binits{X.Y.}},
\bauthor{\bsnm{{Au}}, \binits{B.}},
\bauthor{\bsnm{{Dere}}, \binits{K.P.}},
\bauthor{\bsnm{{Howard}}, \binits{R.A.}},
\bauthor{\bsnm{{Kreplin}}, \binits{R.}},
\bauthor{\bsnm{{Michels}}, \binits{D.J.}},
\bauthor{\bsnm{{Moses}}, \binits{J.D.}},
\bauthor{\bsnm{{Defise}}, \binits{J.M.}},
\bauthor{\bsnm{{Jamar}}, \binits{C.}},
\bauthor{\bsnm{{Rochus}}, \binits{P.}},
\bauthor{\bsnm{{Chauvineau}}, \binits{J.P.}},
\bauthor{\bsnm{{Marioge}}, \binits{J.P.}},
\bauthor{\bsnm{{Catura}}, \binits{R.C.}},
\bauthor{\bsnm{{Lemen}}, \binits{J.R.}},
\bauthor{\bsnm{{Shing}}, \binits{L.}},
\bauthor{\bsnm{{Stern}}, \binits{R.A.}},
\bauthor{\bsnm{{Gurman}}, \binits{J.B.}},
\bauthor{\bsnm{{Neupert}}, \binits{W.M.}},
\bauthor{\bsnm{{Maucherat}}, \binits{A.}},
\bauthor{\bsnm{{Clette}}, \binits{F.}},
\bauthor{\bsnm{{Cugnon}}, \binits{P.}},
\bauthor{\bsnm{{van Dessel}}, \binits{E.L.}}:
\byear{1995},
\batitle{{EIT: Extreme-Ultraviolet Imaging Telescope for the SOHO Mission}}.
\bjtitle{\solphys}
\bvolume{162},
\bfpage{291}.
\doiurl{10.1007/BF00733432}.
\adsurl{1995SoPh..162..291D}.
\end{barticle}
\endbibitem

\bibitem[\protect\citeauthoryear{{Feng} et~al.}{2012}]{feng2012}
\begin{barticle}
\bauthor{\bsnm{{Feng}}, \binits{S.W.}},
\bauthor{\bsnm{{Chen}}, \binits{Y.}},
\bauthor{\bsnm{{Kong}}, \binits{X.L.}},
\bauthor{\bsnm{{Li}}, \binits{G.}},
\bauthor{\bsnm{{Song}}, \binits{H.Q.}},
\bauthor{\bsnm{{Feng}}, \binits{X.S.}},
\bauthor{\bsnm{{Liu}}, \binits{Y.}}:
\byear{2012},
\batitle{{Radio Signatures of Coronal-mass-ejection-Streamer Interaction and
  Source Diagnostics of Type II Radio Burst}}.
\bjtitle{\apj}
\bvolume{753}(\bissue{1}),
\bfpage{21}.
\doiurl{10.1088/0004-637X/753/1/21}.
\adsurl{https://ui.adsabs.harvard.edu/abs/2012ApJ...753...21F}.
\end{barticle}
\endbibitem

\bibitem[\protect\citeauthoryear{{Gopalswamy}}{2006}]{2006GMS...165..207G}
\begin{botherref}
\oauthor{\bsnm{{Gopalswamy}}, \binits{N.}}:
2006,
{Coronal Mass Ejections and Type II Radio Bursts}.
\textit{Geophys. Mono. Ser.165, Am. Geophys. Un., Washington DC, 207}.
\doiurl{10.1029/165GM20}.
\adsurl{https://ui.adsabs.harvard.edu/abs/2006GMS...165..207G}.
\end{botherref}
\endbibitem

\bibitem[\protect\citeauthoryear{{Gopalswamy} and
  {Yashiro}}{2011}]{2011ApJ...736L..17G}
\begin{barticle}
\bauthor{\bsnm{{Gopalswamy}}, \binits{N.}},
\bauthor{\bsnm{{Yashiro}}, \binits{S.}}:
\byear{2011},
\batitle{{The Strength and Radial Profile of the Coronal Magnetic Field from
  the Standoff Distance of a Coronal Mass Ejection-driven Shock}}.
\bjtitle{\apjl}
\bvolume{736}(\bissue{1}),
\bfpage{L17}.
\doiurl{10.1088/2041-8205/736/1/L17}.
\adsurl{https://ui.adsabs.harvard.edu/abs/2011ApJ...736L..17G}.
\end{barticle}
\endbibitem

\bibitem[\protect\citeauthoryear{{Gopalswamy}
  et~al.}{2005}]{2005JGRA..11012S07G}
\begin{barticle}
\bauthor{\bsnm{{Gopalswamy}}, \binits{N.}},
\bauthor{\bsnm{{Aguilar-Rodriguez}}, \binits{E.}},
\bauthor{\bsnm{{Yashiro}}, \binits{S.}},
\bauthor{\bsnm{{Nunes}}, \binits{S.}},
\bauthor{\bsnm{{Kaiser}}, \binits{M.L.}},
\bauthor{\bsnm{{Howard}}, \binits{R.A.}}:
\byear{2005},
\batitle{{Type II radio bursts and energetic solar eruptions}}.
\bjtitle{Geophysi. Res. (Space Physi)}
\bvolume{110}(\bissue{A12}),
\bfpage{A12S07}.
\doiurl{10.1029/2005JA011158}.
\adsurl{https://ui.adsabs.harvard.edu/abs/2005JGRA..11012S07G}.
\end{barticle}
\endbibitem

\bibitem[\protect\citeauthoryear{{Gopalswamy}
  et~al.}{2008}]{2008ApJ...674..560G}
\begin{barticle}
\bauthor{\bsnm{{Gopalswamy}}, \binits{N.}},
\bauthor{\bsnm{{Yashiro}}, \binits{S.}},
\bauthor{\bsnm{{Xie}}, \binits{H.}},
\bauthor{\bsnm{{Akiyama}}, \binits{S.}},
\bauthor{\bsnm{{Aguilar-Rodriguez}}, \binits{E.}},
\bauthor{\bsnm{{Kaiser}}, \binits{M.L.}},
\bauthor{\bsnm{{Howard}}, \binits{R.A.}},
\bauthor{\bsnm{{Bougeret}}, \binits{J.-L.}}:
\byear{2008},
\batitle{{Radio-Quiet Fast and Wide Coronal Mass Ejections}}.
\bjtitle{\apj}
\bvolume{674}(\bissue{1}),
\bfpage{560}.
\doiurl{10.1086/524765}.
\adsurl{https://ui.adsabs.harvard.edu/abs/2008ApJ...674..560G}.
\end{barticle}
\endbibitem

\bibitem[\protect\citeauthoryear{{Gopalswamy}
  et~al.}{2009a}]{2009SoPh..259..227G}
\begin{barticle}
\bauthor{\bsnm{{Gopalswamy}}, \binits{N.}},
\bauthor{\bsnm{{Thompson}}, \binits{W.T.}},
\bauthor{\bsnm{{Davila}}, \binits{J.M.}},
\bauthor{\bsnm{{Kaiser}}, \binits{M.L.}},
\bauthor{\bsnm{{Yashiro}}, \binits{S.}},
\bauthor{\bsnm{{M{\"a}kel{\"a}}}, \binits{P.}},
\bauthor{\bsnm{{Michalek}}, \binits{G.}},
\bauthor{\bsnm{{Bougeret}}, \binits{J.-L.}},
\bauthor{\bsnm{{Howard}}, \binits{R.A.}}:
\byear{2009}a,
\batitle{{Relation Between Type II Bursts and CMEs Inferred from STEREO
  Observations}}.
\bjtitle{\solphys}
\bvolume{259}(\bissue{1-2}),
\bfpage{227}.
\doiurl{10.1007/s11207-009-9382-1}.
\adsurl{https://ui.adsabs.harvard.edu/abs/2009SoPh..259..227G}.
\end{barticle}
\endbibitem

\bibitem[\protect\citeauthoryear{{Gopalswamy}
  et~al.}{2009b}]{Gopalswamy2009EM&PG}
\begin{barticle}
\bauthor{\bsnm{{Gopalswamy}}, \binits{N.}},
\bauthor{\bsnm{{Yashiro}}, \binits{S.}},
\bauthor{\bsnm{{Michalek}}, \binits{G.}},
\bauthor{\bsnm{{Stenborg}}, \binits{G.}},
\bauthor{\bsnm{{Vourlidas}}, \binits{A.}},
\bauthor{\bsnm{{Freeland}}, \binits{S.}},
\bauthor{\bsnm{{Howard}}, \binits{R.}}:
\byear{2009}b,
\batitle{{The SOHO/LASCO CME Catalog}}.
\bjtitle{Earth Moon Planets}
\bvolume{104}(\bissue{1-4}),
\bfpage{295}.
\doiurl{10.1007/s11038-008-9282-7}.
\adsurl{https://ui.adsabs.harvard.edu/abs/2009EM&P..104..295G}.
\end{barticle}
\endbibitem

\bibitem[\protect\citeauthoryear{{Gopalswamy}
  et~al.}{2013}]{2013AdSpR..51.1981G}
\begin{barticle}
\bauthor{\bsnm{{Gopalswamy}}, \binits{N.}},
\bauthor{\bsnm{{Xie}}, \binits{H.}},
\bauthor{\bsnm{{M{\"a}kel{\"a}}}, \binits{P.}},
\bauthor{\bsnm{{Yashiro}}, \binits{S.}},
\bauthor{\bsnm{{Akiyama}}, \binits{S.}},
\bauthor{\bsnm{{Uddin}}, \binits{W.}},
\bauthor{\bsnm{{Srivastava}}, \binits{A.K.}},
\bauthor{\bsnm{{Joshi}}, \binits{N.C.}},
\bauthor{\bsnm{{Chandra}}, \binits{R.}},
\bauthor{\bsnm{{Manoharan}}, \binits{P.K.}},
\bauthor{\bsnm{{Mahalakshmi}}, \binits{K.}},
\bauthor{\bsnm{{Dwivedi}}, \binits{V.C.}},
\bauthor{\bsnm{{Jain}}, \binits{R.}},
\bauthor{\bsnm{{Awasthi}}, \binits{A.K.}},
\bauthor{\bsnm{{Nitta}}, \binits{N.V.}},
\bauthor{\bsnm{{Aschwand en}}, \binits{M.J.}},
\bauthor{\bsnm{{Choudhary}}, \binits{D.P.}}:
\byear{2013},
\batitle{{Height of shock formation in the solar corona inferred from
  observations of type II radio bursts and coronal mass ejections}}.
\bjtitle{Advances in Space Research}
\bvolume{51}(\bissue{11}),
\bfpage{1981}.
\doiurl{10.1016/j.asr.2013.01.006}.
\adsurl{https://ui.adsabs.harvard.edu/abs/2013AdSpR..51.1981G}.
\end{barticle}
\endbibitem

\bibitem[\protect\citeauthoryear{{Gosling}}{1993}]{1993JGR....9818937G}
\begin{barticle}
\bauthor{\bsnm{{Gosling}}, \binits{J.T.}}:
\byear{1993},
\batitle{{The solar flare myth}}.
\bjtitle{\jgr}
\bvolume{98}(\bissue{A11}),
\bfpage{18937}.
\doiurl{10.1029/93JA01896}.
\adsurl{https://ui.adsabs.harvard.edu/abs/1993JGR....9818937G}.
\end{barticle}
\endbibitem

\bibitem[\protect\citeauthoryear{{Gosling} et~al.}{1991}]{1991JGR....96.7831G}
\begin{barticle}
\bauthor{\bsnm{{Gosling}}, \binits{J.T.}},
\bauthor{\bsnm{{McComas}}, \binits{D.J.}},
\bauthor{\bsnm{{Phillips}}, \binits{J.L.}},
\bauthor{\bsnm{{Bame}}, \binits{S.J.}}:
\byear{1991},
\batitle{{Geomagnetic activity associated with earth passage of interplanetary
  shock disturbances and coronal mass ejections}}.
\bjtitle{\jgr}
\bvolume{96}(\bissue{A5}),
\bfpage{7831}.
\doiurl{10.1029/91JA00316}.
\adsurl{https://ui.adsabs.harvard.edu/abs/1991JGR....96.7831G}.
\end{barticle}
\endbibitem

\bibitem[\protect\citeauthoryear{{Holman} and
  {Pesses}}{1983}]{1983ApJ...267..837H}
\begin{barticle}
\bauthor{\bsnm{{Holman}}, \binits{G.D.}},
\bauthor{\bsnm{{Pesses}}, \binits{M.E.}}:
\byear{1983},
\batitle{{Solar type II radio emission and the shock drift acceleration of
  electrons}}.
\bjtitle{\apj}
\bvolume{267},
\bfpage{837}.
\doiurl{10.1086/160918}.
\adsurl{https://ui.adsabs.harvard.edu/abs/1983ApJ...267..837H}.
\end{barticle}
\endbibitem

\bibitem[\protect\citeauthoryear{{Howard} et~al.}{2008}]{2008SSRv..136...67H}
\begin{barticle}
\bauthor{\bsnm{{Howard}}, \binits{R.A.}},
\bauthor{\bsnm{{Moses}}, \binits{J.D.}},
\bauthor{\bsnm{{Vourlidas}}, \binits{A.}},
\bauthor{\bsnm{{Newmark}}, \binits{J.S.}},
\bauthor{\bsnm{{Socker}}, \binits{D.G.}},
\bauthor{\bsnm{{Plunkett}}, \binits{S.P.}},
\bauthor{\bsnm{{Korendyke}}, \binits{C.M.}},
\bauthor{\bsnm{{Cook}}, \binits{J.W.}},
\bauthor{\bsnm{{Hurley}}, \binits{A.}},
\bauthor{\bsnm{{Davila}}, \binits{J.M.}},
\bauthor{\bsnm{{Thompson}}, \binits{W.T.}},
\bauthor{\bsnm{{St Cyr}}, \binits{O.C.}},
\bauthor{\bsnm{{Mentzell}}, \binits{E.}},
\bauthor{\bsnm{{Mehalick}}, \binits{K.}},
\bauthor{\bsnm{{Lemen}}, \binits{J.R.}},
\bauthor{\bsnm{{Wuelser}}, \binits{J.P.}},
\bauthor{\bsnm{{Duncan}}, \binits{D.W.}},
\bauthor{\bsnm{{Tarbell}}, \binits{T.D.}},
\bauthor{\bsnm{{Wolfson}}, \binits{C.J.}},
\bauthor{\bsnm{{Moore}}, \binits{A.}},
\bauthor{\bsnm{{Harrison}}, \binits{R.A.}},
\bauthor{\bsnm{{Waltham}}, \binits{N.R.}},
\bauthor{\bsnm{{Lang}}, \binits{J.}},
\bauthor{\bsnm{{Davis}}, \binits{C.J.}},
\bauthor{\bsnm{{Eyles}}, \binits{C.J.}},
\bauthor{\bsnm{{Mapson-Menard}}, \binits{H.}},
\bauthor{\bsnm{{Simnett}}, \binits{G.M.}},
\bauthor{\bsnm{{Halain}}, \binits{J.P.}},
\bauthor{\bsnm{{Defise}}, \binits{J.M.}},
\bauthor{\bsnm{{Mazy}}, \binits{E.}},
\bauthor{\bsnm{{Rochus}}, \binits{P.}},
\bauthor{\bsnm{{Mercier}}, \binits{R.}},
\bauthor{\bsnm{{Ravet}}, \binits{M.F.}},
\bauthor{\bsnm{{Delmotte}}, \binits{F.}},
\bauthor{\bsnm{{Auchere}}, \binits{F.}},
\bauthor{\bsnm{{Delaboudini\'ere}}, \binits{J.P.}},
\bauthor{\bsnm{{Bothmer}}, \binits{V.}},
\bauthor{\bsnm{{Deutsch}}, \binits{W.}},
\bauthor{\bsnm{{Wang}}, \binits{D.}},
\bauthor{\bsnm{{Rich}}, \binits{N.}},
\bauthor{\bsnm{{Cooper}}, \binits{S.}},
\bauthor{\bsnm{{Stephens}}, \binits{V.}},
\bauthor{\bsnm{{Maahs}}, \binits{G.}},
\bauthor{\bsnm{{Baugh}}, \binits{R.}},
\bauthor{\bsnm{{McMullin}}, \binits{D.}},
\bauthor{\bsnm{{Carter}}, \binits{T.}}:
\byear{2008},
\batitle{{Sun Earth Connection Coronal and Heliospheric Investigation
  (SECCHI)}}.
\bjtitle{\ssr}
\bvolume{136}(\bissue{1-4}),
\bfpage{67}.
\doiurl{10.1007/s11214-008-9341-4}.
\adsurl{https://ui.adsabs.harvard.edu/abs/2008SSRv..136...67H}.
\end{barticle}
\endbibitem

\bibitem[\protect\citeauthoryear{{Hundhausen}}{1987}]{1987sowi.conf..181H}
\begin{bchapter}
\bauthor{\bsnm{{Hundhausen}}, \binits{A.J.}}:
\byear{1987},
\bctitle{{The Origin and Propagation of Coronal Mass Ejections (R)}}.
In: \beditor{\bsnm{{Pizzo}}, \binits{V.J.}},
\beditor{\bsnm{{Holzer}}, \binits{T.}},
\beditor{\bsnm{{Sime}}, \binits{D.G.}} (eds.)
\bbtitle{Sixth International Solar Wind Conference}
\bseriesno{2},
\bfpage{181}.
\adsurl{https://ui.adsabs.harvard.edu/abs/1987sowi.conf..181H}.
\end{bchapter}
\endbibitem

\bibitem[\protect\citeauthoryear{{Jebaraj} et~al.}{2020}]{2020A&A...639A..56J}
\begin{barticle}
\bauthor{\bsnm{{Jebaraj}}, \binits{I.C.}},
\bauthor{\bsnm{{Magdaleni{\'c}}}, \binits{J.}},
\bauthor{\bsnm{{Podladchikova}}, \binits{T.}},
\bauthor{\bsnm{{Scolini}}, \binits{C.}},
\bauthor{\bsnm{{Pomoell}}, \binits{J.}},
\bauthor{\bsnm{{Veronig}}, \binits{A.M.}},
\bauthor{\bsnm{{Dissauer}}, \binits{K.}},
\bauthor{\bsnm{{Krupar}}, \binits{V.}},
\bauthor{\bsnm{{Kilpua}}, \binits{E.K.J.}},
\bauthor{\bsnm{{Poedts}}, \binits{S.}}:
\byear{2020},
\batitle{{Using radio triangulation to understand the origin of two subsequent
  type II radio bursts}}.
\bjtitle{\aap}
\bvolume{639},
\bfpage{A56}.
\doiurl{10.1051/0004-6361/201937273}.
\adsurl{https://ui.adsabs.harvard.edu/abs/2020A&A...639A..56J}.
\end{barticle}
\endbibitem

\bibitem[\protect\citeauthoryear{{Kahler}, {Hildner}, and {Van
  Hollebeke}}{1978}]{1978SoPh...57..429K}
\begin{barticle}
\bauthor{\bsnm{{Kahler}}, \binits{S.W.}},
\bauthor{\bsnm{{Hildner}}, \binits{E.}},
\bauthor{\bsnm{{Van Hollebeke}}, \binits{M.A.I.}}:
\byear{1978},
\batitle{{Promt solar proton events and coronal mass ejections.}}
\bjtitle{\solphys}
\bvolume{57}(\bissue{2}),
\bfpage{429}.
\doiurl{10.1007/BF00160116}.
\adsurl{https://ui.adsabs.harvard.edu/abs/1978SoPh...57..429K}.
\end{barticle}
\endbibitem

\bibitem[\protect\citeauthoryear{{Kahler}, {Ling}, and
  {Gopalswamy}}{2019}]{Kahler2019}
\begin{barticle}
\bauthor{\bsnm{{Kahler}}, \binits{S.W.}},
\bauthor{\bsnm{{Ling}}, \binits{A.G.}},
\bauthor{\bsnm{{Gopalswamy}}, \binits{N.}}:
\byear{2019},
\batitle{{Are Solar Energetic Particle Events and Type II Bursts Associated
  with Fast and Narrow Coronal Mass Ejections?}}
\bjtitle{\solphys}
\bvolume{294}(\bissue{9}),
\bfpage{134}.
\doiurl{10.1007/s11207-019-1518-3}.
\adsurl{https://ui.adsabs.harvard.edu/abs/2019SoPh..294..134K}.
\end{barticle}
\endbibitem

\bibitem[\protect\citeauthoryear{{Kaiser} et~al.}{2008}]{2008SSRv..136....5K}
\begin{barticle}
\bauthor{\bsnm{{Kaiser}}, \binits{M.L.}},
\bauthor{\bsnm{{Kucera}}, \binits{T.A.}},
\bauthor{\bsnm{{Davila}}, \binits{J.M.}},
\bauthor{\bsnm{{St. Cyr}}, \binits{O.C.}},
\bauthor{\bsnm{{Guhathakurta}}, \binits{M.}},
\bauthor{\bsnm{{Christian}}, \binits{E.}}:
\byear{2008},
\batitle{{The STEREO Mission: An Introduction}}.
\bjtitle{\ssr}
\bvolume{136}(\bissue{1-4}),
\bfpage{5}.
\doiurl{10.1007/s11214-007-9277-0}.
\adsurl{https://ui.adsabs.harvard.edu/abs/2008SSRv..136....5K}.
\end{barticle}
\endbibitem

\bibitem[\protect\citeauthoryear{{Kihara} et~al.}{2020}]{2020ApJ...900...75K}
\begin{barticle}
\bauthor{\bsnm{{Kihara}}, \binits{K.}},
\bauthor{\bsnm{{Huang}}, \binits{Y.}},
\bauthor{\bsnm{{Nishimura}}, \binits{N.}},
\bauthor{\bsnm{{Nitta}}, \binits{N.V.}},
\bauthor{\bsnm{{Yashiro}}, \binits{S.}},
\bauthor{\bsnm{{Ichimoto}}, \binits{K.}},
\bauthor{\bsnm{{Asai}}, \binits{A.}}:
\byear{2020},
\batitle{{Statistical Analysis of the Relation between Coronal Mass Ejections
  and Solar Energetic Particles}}.
\bjtitle{\apj}
\bvolume{900}(\bissue{1}),
\bfpage{75}.
\doiurl{10.3847/1538-4357/aba621}.
\adsurl{https://ui.adsabs.harvard.edu/abs/2020ApJ...900...75K}.
\end{barticle}
\endbibitem

\bibitem[\protect\citeauthoryear{{Kilpua} et~al.}{2013}]{2013AnGeo..31.1251K}
\begin{barticle}
\bauthor{\bsnm{{Kilpua}}, \binits{E.K.J.}},
\bauthor{\bsnm{{Isavnin}}, \binits{A.}},
\bauthor{\bsnm{{Vourlidas}}, \binits{A.}},
\bauthor{\bsnm{{Koskinen}}, \binits{H.E.J.}},
\bauthor{\bsnm{{Rodriguez}}, \binits{L.}}:
\byear{2013},
\batitle{{On the relationship between interplanetary coronal mass ejections and
  magnetic clouds}}.
\bjtitle{Annales Geophysicae}
\bvolume{31}(\bissue{7}),
\bfpage{1251}.
\doiurl{10.5194/angeo-31-1251-2013}.
\adsurl{https://ui.adsabs.harvard.edu/abs/2013AnGeo..31.1251K}.
\end{barticle}
\endbibitem

\bibitem[\protect\citeauthoryear{{Kishore} et~al.}{2014}]{2014SoPh..289.3995K}
\begin{barticle}
\bauthor{\bsnm{{Kishore}}, \binits{P.}},
\bauthor{\bsnm{{Kathiravan}}, \binits{C.}},
\bauthor{\bsnm{{Ramesh}}, \binits{R.}},
\bauthor{\bsnm{{Rajalingam}}, \binits{M.}},
\bauthor{\bsnm{{Barve}}, \binits{I.V.}}:
\byear{2014},
\batitle{{Gauribidanur Low-Frequency Solar Spectrograph}}.
\bjtitle{\solphys}
\bvolume{289}(\bissue{10}),
\bfpage{3995}.
\doiurl{10.1007/s11207-014-0539-1}.
\adsurl{https://ui.adsabs.harvard.edu/abs/2014SoPh..289.3995K}.
\end{barticle}
\endbibitem

\bibitem[\protect\citeauthoryear{{Kumari} et~al.}{2017a}]{kumari2017c}
\begin{barticle}
\bauthor{\bsnm{{Kumari}}, \binits{A.}},
\bauthor{\bsnm{{Ramesh}}, \binits{R.}},
\bauthor{\bsnm{{Kathiravan}}, \binits{C.}},
\bauthor{\bsnm{{Wang}}, \binits{T.J.}}:
\byear{2017}a,
\batitle{{Addendum to: Strength of the Solar Coronal Magnetic Field - A
  Comparison of Independent Estimates Using Contemporaneous Radio and
  White-Light Observations}}.
\bjtitle{Solar Phys.}
\bvolume{292}(\bissue{12}),
\bfpage{177}.
\doiurl{10.1007/s11207-017-1203-3}.
\adsurl{https://ui.adsabs.harvard.edu/abs/2017SoPh..292..177K}.
\end{barticle}
\endbibitem

\bibitem[\protect\citeauthoryear{{Kumari} et~al.}{2017b}]{kumari2017a}
\begin{barticle}
\bauthor{\bsnm{{Kumari}}, \binits{A.}},
\bauthor{\bsnm{{Ramesh}}, \binits{R.}},
\bauthor{\bsnm{{Kathiravan}}, \binits{C.}},
\bauthor{\bsnm{{Gopalswamy}}, \binits{N.}}:
\byear{2017}b,
\batitle{{New Evidence for a Coronal Mass Ejection-driven High Frequency Type
  II Burst near the Sun}}.
\bjtitle{\apj}
\bvolume{843},
\bfpage{10}.
\doiurl{10.3847/1538-4357/aa72e7}.
\adsurl{2017ApJ...843...10K}.
\end{barticle}
\endbibitem

\bibitem[\protect\citeauthoryear{{Kumari} et~al.}{2017c}]{kumari2017b}
\begin{barticle}
\bauthor{\bsnm{{Kumari}}, \binits{A.}},
\bauthor{\bsnm{{Ramesh}}, \binits{R.}},
\bauthor{\bsnm{{Kathiravan}}, \binits{C.}},
\bauthor{\bsnm{{Wang}}, \binits{T.J.}}:
\byear{2017}c,
\batitle{{Strength of the Solar Coronal Magnetic Field - A Comparison of
  Independent Estimates Using Contemporaneous Radio and White-Light
  Observations}}.
\bjtitle{Solar Phys.}
\bvolume{292}(\bissue{11}),
\bfpage{161}.
\doiurl{10.1007/s11207-017-1180-6}.
\adsurl{https://ui.adsabs.harvard.edu/abs/2017SoPh..292..161K}.
\end{barticle}
\endbibitem

\bibitem[\protect\citeauthoryear{Kumari et~al.}{2019}]{kumari2019}
\begin{barticle}
\bauthor{\bsnm{Kumari}, \binits{A.}},
\bauthor{\bsnm{Ramesh}, \binits{R.}},
\bauthor{\bsnm{Kathiravan}, \binits{C.}},
\bauthor{\bsnm{Wang}, \binits{T.J.}},
\bauthor{\bsnm{Gopalswamy}, \binits{N.}}:
\byear{2019},
\batitle{Direct estimates of the solar coronal magnetic field using
  contemporaneous extreme-ultraviolet, radio, and white-light observations}.
\bjtitle{Astrophys. J.}
\bvolume{881}(\bissue{1}),
\bfpage{24}.
\doiurl{10.3847/1538-4357/ab2adf}.
\end{barticle}
\endbibitem

\bibitem[\protect\citeauthoryear{Lemen et~al.}{2011}]{aia}
\begin{barticle}
\bauthor{\bsnm{Lemen}, \binits{J.}},
\bauthor{\bsnm{Title}, \binits{A.}},
\bauthor{\bsnm{Boerner}, \binits{P.}},
\bauthor{\bsnm{Chou}, \binits{C.}},
\bauthor{\bsnm{Drake}, \binits{J.}},
\bauthor{\bsnm{Duncan}, \binits{D.}},
\bauthor{\bsnm{Edwards}, \binits{C.}},
\bauthor{\bsnm{Friedlaender}, \binits{F.}},
\bauthor{\bsnm{Heyman}, \binits{G.}},
\bauthor{\bsnm{Hurlburt}, \binits{N.}},
\bauthor{\bsnm{Katz}, \binits{N.}},
\bauthor{\bsnm{Kushner}, \binits{G.}},
\bauthor{\bsnm{Levay}, \binits{M.}},
\bauthor{\bsnm{Lindgren}, \binits{R.}},
\bauthor{\bsnm{Mathur}, \binits{D.}},
\bauthor{\bsnm{McFeaters}, \binits{E.}},
\bauthor{\bsnm{Mitchell}, \binits{S.}},
\bauthor{\bsnm{Rehse}, \binits{R.}},
\bauthor{\bsnm{Waltham}, \binits{N.}}:
\byear{2011},
\batitle{The atmospheric imaging assembly (aia) on the solar dynamics
  observatory (sdo)}.
\bjtitle{Solar Phys.}
\bvolume{275},
\bfpage{17}.
\doiurl{10.1007/s11207-011-9776-8}.
\end{barticle}
\endbibitem

\bibitem[\protect\citeauthoryear{{Monstein}}{2013}]{2013EGUGA..15.2027M}
\begin{bchapter}
\bauthor{\bsnm{{Monstein}}, \binits{C.}}:
\byear{2013},
\bctitle{{CALLISTO and the e-CALLISTO network}}.
In: \bbtitle{EGU Gen Assemb Conf Abs}.
\adsurl{https://ui.adsabs.harvard.edu/abs/2013EGUGA..15.2027M}.
\end{bchapter}
\endbibitem

\bibitem[\protect\citeauthoryear{Morosan et~al.}{2020}]{morosan2020electron}
\begin{barticle}
\bauthor{\bsnm{Morosan}, \binits{D.}},
\bauthor{\bsnm{R{\"a}s{\"a}nen}, \binits{E.}},
\bauthor{\bsnm{Kilpua}, \binits{E.}},
\bauthor{\bsnm{Magdaleni{\'c}}, \binits{J.}},
\bauthor{\bsnm{Lynch}, \binits{B.}},
\bauthor{\bsnm{Kumari}, \binits{A.}},
\bauthor{\bsnm{Pomoell}, \binits{J.}},
\bauthor{\bsnm{Palmroth}, \binits{M.}}:
\byear{2020},
\batitle{{Electron acceleration and radio emission following the early
  interaction of two coronal mass ejections}}.
\bjtitle{\aap}
\bvolume{A151}(\bissue{9}),
\bfpage{134}.
\doiurl{10.1051/0004-6361/202038801}.
\adsurl{https://ui.adsabs.harvard.edu/abs/2020A\%26A...642A.151M}.
\end{barticle}
\endbibitem

\bibitem[\protect\citeauthoryear{M{\"u}ller et~al.}{2013}]{solarorbiter}
\begin{barticle}
\bauthor{\bsnm{M{\"u}ller}, \binits{D.}},
\bauthor{\bsnm{Marsden}, \binits{R.G.}},
\bauthor{\bsnm{St.~Cyr}, \binits{O.C.}},
\bauthor{\bsnm{Gilbert}, \binits{H.R.}},
\bauthor{\bsnm{{The Solar Orbiter Team}}}:
\byear{2013},
\batitle{Solar orbiter}.
\bjtitle{Solar Phys.}
\bvolume{285}(\bissue{1}),
\bfpage{25}.
\doiurl{10.1007/s11207-012-0085-7}.
\end{barticle}
\endbibitem

\bibitem[\protect\citeauthoryear{{Newkirk}}{1967}]{newkirk1967}
\begin{barticle}
\bauthor{\bsnm{{Newkirk}}, \binits{J.} \bsuffix{Gordon}}:
\byear{1967},
\batitle{{Structure of the Solar Corona}}.
\bjtitle{Annu. Rev. Astron. Astrophys}
\bvolume{5},
\bfpage{213}.
\doiurl{10.1146/annurev.aa.05.090167.001241}.
\adsurl{https://ui.adsabs.harvard.edu/abs/1967ARA&A...5..213N}.
\end{barticle}
\endbibitem

\bibitem[\protect\citeauthoryear{{Pick} et~al.}{2006}]{2006SSRv..123..341P}
\begin{barticle}
\bauthor{\bsnm{{Pick}}, \binits{M.}},
\bauthor{\bsnm{{Forbes}}, \binits{T.G.}},
\bauthor{\bsnm{{Mann}}, \binits{G.}},
\bauthor{\bsnm{{Cane}}, \binits{H.V.}},
\bauthor{\bsnm{{Chen}}, \binits{J.}},
\bauthor{\bsnm{{Ciaravella}}, \binits{A.}},
\bauthor{\bsnm{{Cremades}}, \binits{H.}},
\bauthor{\bsnm{{Howard}}, \binits{R.A.}},
\bauthor{\bsnm{{Hudson}}, \binits{H.S.}},
\bauthor{\bsnm{{Klassen}}, \binits{A.}},
\bauthor{\bsnm{{Klein}}, \binits{K.L.}},
\bauthor{\bsnm{{Lee}}, \binits{M.A.}},
\bauthor{\bsnm{{Linker}}, \binits{J.A.}},
\bauthor{\bsnm{{Maia}}, \binits{D.}},
\bauthor{\bsnm{{Mikic}}, \binits{Z.}},
\bauthor{\bsnm{{Raymond}}, \binits{J.C.}},
\bauthor{\bsnm{{Reiner}}, \binits{M.J.}},
\bauthor{\bsnm{{Simnett}}, \binits{G.M.}},
\bauthor{\bsnm{{Srivastava}}, \binits{N.}},
\bauthor{\bsnm{{Tripathi}}, \binits{D.}},
\bauthor{\bsnm{{Vainio}}, \binits{R.}},
\bauthor{\bsnm{{Vourlidas}}, \binits{A.}},
\bauthor{\bsnm{{Zhang}}, \binits{J.}},
\bauthor{\bsnm{{Zurbuchen}}, \binits{T.H.}},
\bauthor{\bsnm{{Sheeley}}, \binits{N.R.}},
\bauthor{\bsnm{{Marqu{\'e}}}, \binits{C.}}:
\byear{2006},
\batitle{{Multi-Wavelength Observations of CMEs and Associated Phenomena.
  Report of Working Group F}}.
\bjtitle{\ssr}
\bvolume{123}(\bissue{1-3}),
\bfpage{341}.
\doiurl{10.1007/s11214-006-9021-1}.
\adsurl{https://ui.adsabs.harvard.edu/abs/2006SSRv..123..341P}.
\end{barticle}
\endbibitem

\bibitem[\protect\citeauthoryear{{Raghavendra Prasad} et~al.}{2017}]{VELC17}
\begin{barticle}
\bauthor{\bsnm{{Raghavendra Prasad}}, \binits{B.}},
\bauthor{\bsnm{Banerjee}, \binits{D.}},
\bauthor{\bsnm{Singh}, \binits{J.}},
\bauthor{\bsnm{Nagabhushana}, \binits{S.}},
\bauthor{\bsnm{Kumar}, \binits{A.}},
\bauthor{\bsnm{Kamath}, \binits{P.U.}},
\bauthor{\bsnm{Kathiravan}, \binits{S.}},
\bauthor{\bsnm{Venkata}, \binits{S.}},
\bauthor{\bsnm{Rajkumar}, \binits{N.}},
\bauthor{\bsnm{Natarajan}, \binits{V.}},
\bauthor{\bsnm{Juneja}, \binits{M.}},
\bauthor{\bsnm{Somu}, \binits{P.}},
\bauthor{\bsnm{Pant}, \binits{V.}},
\bauthor{\bsnm{Shaji}, \binits{N.}},
\bauthor{\bsnm{Sankarsubramanian}, \binits{K.}},
\bauthor{\bsnm{Patra}, \binits{A.}},
\bauthor{\bsnm{Venkateswaran}, \binits{R.}},
\bauthor{\bsnm{Adoni}, \binits{A.A.}},
\bauthor{\bsnm{Narendra}, \binits{S.}},
\bauthor{\bsnm{Haridas}, \binits{T.R.}},
\bauthor{\bsnm{Mathew}, \binits{S.K.}},
\bauthor{\bsnm{Krishna}, \binits{R.M.}},
\bauthor{\bsnm{Amareswari}, \binits{K.}},
\bauthor{\bsnm{Jaiswal}, \binits{B.}}:
\byear{2017},
\batitle{{Visible Emission Line Coronagraph on Aditya-L1}}.
\bjtitle{Current Science}
\bvolume{113}(\bissue{4}),
\bfpage{613}.
\doiurl{10.18520/cs/v113/i04/613-615}.
\end{barticle}
\endbibitem

\bibitem[\protect\citeauthoryear{{Ramesh} et~al.}{1998}]{1998SoPh..181..439R}
\begin{barticle}
\bauthor{\bsnm{{Ramesh}}, \binits{R.}},
\bauthor{\bsnm{{Subramanian}}, \binits{K.R.}},
\bauthor{\bsnm{{Sundararajan}}, \binits{M.S.}},
\bauthor{\bsnm{{Sastry}}, \binits{C.V.}}:
\byear{1998},
\batitle{{The Gauribidanur Radioheliograph}}.
\bjtitle{\solphys}
\bvolume{181}(\bissue{2}),
\bfpage{439}.
\doiurl{10.1023/A:1005075003370}.
\adsurl{https://ui.adsabs.harvard.edu/abs/1998SoPh..181..439R}.
\end{barticle}
\endbibitem

\bibitem[\protect\citeauthoryear{{Ramesh} et~al.}{2010}]{2010ApJ...712..188R}
\begin{barticle}
\bauthor{\bsnm{{Ramesh}}, \binits{R.}},
\bauthor{\bsnm{{Kathiravan}}, \binits{C.}},
\bauthor{\bsnm{{Kartha}}, \binits{S.S.}},
\bauthor{\bsnm{{Gopalswamy}}, \binits{N.}}:
\byear{2010},
\batitle{{Radioheliograph Observations of Metric Type II Bursts and the
  Kinematics of Coronal Mass Ejections}}.
\bjtitle{\apj}
\bvolume{712}(\bissue{1}),
\bfpage{188}.
\doiurl{10.1088/0004-637X/712/1/188}.
\adsurl{https://ui.adsabs.harvard.edu/abs/2010ApJ...712..188R}.
\end{barticle}
\endbibitem

\bibitem[\protect\citeauthoryear{{Reames}}{1999}]{1999SSRv...90..413R}
\begin{barticle}
\bauthor{\bsnm{{Reames}}, \binits{D.V.}}:
\byear{1999},
\batitle{{Particle acceleration at the Sun and in the heliosphere}}.
\bjtitle{\ssr}
\bvolume{90},
\bfpage{413}.
\doiurl{10.1023/A:1005105831781}.
\adsurl{https://ui.adsabs.harvard.edu/abs/1999SSRv...90..413R}.
\end{barticle}
\endbibitem

\bibitem[\protect\citeauthoryear{{Renotte} et~al.}{2014}]{Proba3}
\begin{bchapter}
\bauthor{\bsnm{{Renotte}}, \binits{E.}},
\bauthor{\bsnm{{Baston}}, \binits{E.C.}},
\bauthor{\bsnm{{Bemporad}}, \binits{A.}},
\bauthor{\bsnm{{Capobianco}}, \binits{G.}},
\bauthor{\bsnm{{Cernica}}, \binits{I.}},
\bauthor{\bsnm{{Darakchiev}}, \binits{R.}},
\bauthor{\bsnm{{Denis}}, \binits{F.}},
\bauthor{\bsnm{{Desselle}}, \binits{R.}},
\bauthor{\bsnm{{De Vos}}, \binits{L.}},
\bauthor{\bsnm{{Fineschi}}, \binits{S.}},
\bauthor{\bsnm{{Focardi}}, \binits{M.}},
\bauthor{\bsnm{{G{\'o}rski}}, \binits{T.}},
\bauthor{\bsnm{{Graczyk}}, \binits{R.}},
\bauthor{\bsnm{{Halain}}, \binits{J.-P.}},
\bauthor{\bsnm{{Hermans}}, \binits{A.}},
\bauthor{\bsnm{{Jackson}}, \binits{C.}},
\bauthor{\bsnm{{Kintziger}}, \binits{C.}},
\bauthor{\bsnm{{Kosiec}}, \binits{J.}},
\bauthor{\bsnm{{Kranitis}}, \binits{N.}},
\bauthor{\bsnm{{Landini}}, \binits{F.}},
\bauthor{\bsnm{{L{\'e}dl}}, \binits{V.}},
\bauthor{\bsnm{{Massone}}, \binits{G.}},
\bauthor{\bsnm{{Mazzoli}}, \binits{A.}},
\bauthor{\bsnm{{Melich}}, \binits{R.}},
\bauthor{\bsnm{{Mollet}}, \binits{D.}},
\bauthor{\bsnm{{Mosdorf}}, \binits{M.}},
\bauthor{\bsnm{{Nicolini}}, \binits{G.}},
\bauthor{\bsnm{{Nicula}}, \binits{B.}},
\bauthor{\bsnm{{Orlea{\'n}ski}}, \binits{P.}},
\bauthor{\bsnm{{Palau}}, \binits{M.-C.}},
\bauthor{\bsnm{{Pancrazzi}}, \binits{M.}},
\bauthor{\bsnm{{Paschalis}}, \binits{A.}},
\bauthor{\bsnm{{Peresty}}, \binits{R.}},
\bauthor{\bsnm{{Plesseria}}, \binits{J.-Y.}},
\bauthor{\bsnm{{Rataj}}, \binits{M.}},
\bauthor{\bsnm{{Romoli}}, \binits{M.}},
\bauthor{\bsnm{{Thizy}}, \binits{C.}},
\bauthor{\bsnm{{Thom{\'e}}}, \binits{M.}},
\bauthor{\bsnm{{Tsinganos}}, \binits{K.}},
\bauthor{\bsnm{{Wodnicki}}, \binits{R.}},
\bauthor{\bsnm{{Walczak}}, \binits{T.}},
\bauthor{\bsnm{{Zhukov}}, \binits{A.}}:
\byear{2014},
\bctitle{{ASPIICS: an externally occulted coronagraph for PROBA-3: Design
  evolution}}.
In: \beditor{\bsnm{{Oschmann}}, \binits{J.} \bsuffix{Jacobus~M.}},
\beditor{\bsnm{{Clampin}}, \binits{M.}},
\beditor{\bsnm{{Fazio}}, \binits{G.G.}},
\beditor{\bsnm{{MacEwen}}, \binits{H.A.}} (eds.)
\bbtitle{Space Telescopes and Instrumentation 2014: Optical, Infrared, and
  Millimeter Wave},
\bsertitle{Pro. SPIE}
\bseriesno{9143},
\bfpage{91432M}.
\doiurl{10.1117/12.2056784}.
\adsurl{2014SPIE.9143E..2MR}.
\end{bchapter}
\endbibitem

\bibitem[\protect\citeauthoryear{{Richardson} and
  {Cane}}{2012}]{2012JSWSC...2A..01R}
\begin{barticle}
\bauthor{\bsnm{{Richardson}}, \binits{I.G.}},
\bauthor{\bsnm{{Cane}}, \binits{H.V.}}:
\byear{2012},
\batitle{{Solar wind drivers of geomagnetic storms during more than four solar
  cycles}}.
\bjtitle{Space Weather Space Climate}
\bvolume{2},
\bfpage{A01}.
\doiurl{10.1051/swsc/2012001}.
\adsurl{https://ui.adsabs.harvard.edu/abs/2012JSWSC...2A..01R}.
\end{barticle}
\endbibitem

\bibitem[\protect\citeauthoryear{Russell and Mulligan}{2002}]{RUSSELL2002527}
\begin{barticle}
\bauthor{\bsnm{Russell}, \binits{C.T.}},
\bauthor{\bsnm{Mulligan}, \binits{T.}}:
\byear{2002},
\batitle{On the magnetosheath thicknesses of interplanetary coronal mass
  ejections}.
\bjtitle{Planet Space Sci}
\bvolume{50}(\bissue{5}),
\bfpage{527 }.
\doiurl{https://doi.org/10.1016/S0032-0633(02)00031-4}.
\end{barticle}
\endbibitem

\bibitem[\protect\citeauthoryear{Seetha and Megala}{2017}]{ADITYA2017}
\begin{barticle}
\bauthor{\bsnm{Seetha}, \binits{S.}},
\bauthor{\bsnm{Megala}, \binits{S.}}:
\byear{2017},
\batitle{{Aditya-L1 mission}}.
\bjtitle{Cur. Science}
\bvolume{113}(\bissue{4}),
\bfpage{610}.
\doiurl{10.18520/cs/v113/i04/ 610 - 612}.
\end{barticle}
\endbibitem

\bibitem[\protect\citeauthoryear{{Sheeley}, {Hakala}, and
  {Wang}}{2000}]{2000JGR...105.5081S}
\begin{barticle}
\bauthor{\bsnm{{Sheeley}}, \binits{N.R.}},
\bauthor{\bsnm{{Hakala}}, \binits{W.N.}},
\bauthor{\bsnm{{Wang}}, \binits{Y.-M.}}:
\byear{2000},
\batitle{{Detection of coronal mass ejection associated shock waves in the
  outer corona}}.
\bjtitle{\jgr}
\bvolume{105}(\bissue{A3}),
\bfpage{5081}.
\doiurl{10.1029/1999JA000338}.
\adsurl{https://ui.adsabs.harvard.edu/abs/2000JGR...105.5081S}.
\end{barticle}
\endbibitem

\bibitem[\protect\citeauthoryear{{Smerd}, {Sheridan}, and
  {Stewart}}{1974}]{1974IAUS...57..389S}
\begin{bchapter}
\bauthor{\bsnm{{Smerd}}, \binits{S.F.}},
\bauthor{\bsnm{{Sheridan}}, \binits{K.V.}},
\bauthor{\bsnm{{Stewart}}, \binits{R.T.}}:
\byear{1974},
\bctitle{{On Split-Band Structure in Type II Radio Bursts from the Sun}}.
In: \beditor{\bsnm{{Newkirk}}, \binits{G.A.}} (ed.)
\bbtitle{Coronal Disturbances},
\bsertitle{Pro. IAU Sym}
\bseriesno{57},
\bfpage{389}.
\adsurl{https://ui.adsabs.harvard.edu/abs/1974IAUS...57..389S}.
\end{bchapter}
\endbibitem

\bibitem[\protect\citeauthoryear{{Thernisien}, {Vourlidas}, and
  {Howard}}{2009}]{2009SoPh..256..111T}
\begin{barticle}
\bauthor{\bsnm{{Thernisien}}, \binits{A.}},
\bauthor{\bsnm{{Vourlidas}}, \binits{A.}},
\bauthor{\bsnm{{Howard}}, \binits{R.A.}}:
\byear{2009},
\batitle{{Forward Modeling of Coronal Mass Ejections Using STEREO/SECCHI
  Data}}.
\bjtitle{\solphys}
\bvolume{256}(\bissue{1-2}),
\bfpage{111}.
\doiurl{10.1007/s11207-009-9346-5}.
\adsurl{https://ui.adsabs.harvard.edu/abs/2009SoPh..256..111T}.
\end{barticle}
\endbibitem

\bibitem[\protect\citeauthoryear{Vourlidas et~al.}{2008}]{vourlidas2008}
\begin{barticle}
\bauthor{\bsnm{Vourlidas}, \binits{A.}},
\bauthor{\bsnm{Wu}, \binits{S.}},
\bauthor{\bsnm{Wang}, \binits{A.H.}},
\bauthor{\bsnm{Subramanian}, \binits{P.}},
\bauthor{\bsnm{Howard}, \binits{R.}}:
\byear{2008},
\batitle{Direct detection of a coronal mass ejection-associated shock in large
  angle and spectrometric coronagraph experiment white-light images}.
\bjtitle{Astrophys. J.}
\bvolume{598},
\bfpage{1392}.
\doiurl{10.1086/379098}.
\end{barticle}
\endbibitem

\bibitem[\protect\citeauthoryear{{Vr{\v{s}}nak}, {Magdaleni{\'c}}, and
  {Zlobec}}{2004}]{vrsnak_2004}
\begin{barticle}
\bauthor{\bsnm{{Vr{\v{s}}nak}}, \binits{B.}},
\bauthor{\bsnm{{Magdaleni{\'c}}}, \binits{J.}},
\bauthor{\bsnm{{Zlobec}}, \binits{P.}}:
\byear{2004},
\batitle{{Band-splitting of coronal and interplanetary type II bursts. III.
  Physical conditions in the upper corona and interplanetary space}}.
\bjtitle{\aap}
\bvolume{413},
\bfpage{753}.
\doiurl{10.1051/0004-6361:20034060}.
\adsurl{https://ui.adsabs.harvard.edu/abs/2004A&A...413..753V}.
\end{barticle}
\endbibitem

\bibitem[\protect\citeauthoryear{{Vr{\v{s}}nak} et~al.}{2001}]{vrsnak_2001}
\begin{barticle}
\bauthor{\bsnm{{Vr{\v{s}}nak}}, \binits{B.}},
\bauthor{\bsnm{{Aurass}}, \binits{H.}},
\bauthor{\bsnm{{Magdaleni{\'c}}}, \binits{J.}},
\bauthor{\bsnm{{Gopalswamy}}, \binits{N.}}:
\byear{2001},
\batitle{{Band-splitting of coronal and interplanetary type II bursts. I. Basic
  properties}}.
\bjtitle{\aap}
\bvolume{377},
\bfpage{321}.
\doiurl{10.1051/0004-6361:20011067}.
\adsurl{https://ui.adsabs.harvard.edu/abs/2001A&A...377..321V}.
\end{barticle}
\endbibitem

\bibitem[\protect\citeauthoryear{{Vr{\v{s}}nak} et~al.}{2002}]{vrsnak_2002}
\begin{barticle}
\bauthor{\bsnm{{Vr{\v{s}}nak}}, \binits{B.}},
\bauthor{\bsnm{{Magdaleni{\'c}}}, \binits{J.}},
\bauthor{\bsnm{{Aurass}}, \binits{H.}},
\bauthor{\bsnm{{Mann}}, \binits{G.}}:
\byear{2002},
\batitle{{Band-splitting of coronal and interplanetary type II bursts. II.
  Coronal magnetic field and Alfv{\'e}n velocity}}.
\bjtitle{\aap}
\bvolume{396},
\bfpage{673}.
\doiurl{10.1051/0004-6361:20021413}.
\adsurl{https://ui.adsabs.harvard.edu/abs/2002A&A...396..673V}.
\end{barticle}
\endbibitem

\bibitem[\protect\citeauthoryear{{Vr{\v{s}}nak} et~al.}{2006}]{Vrsnak2006}
\begin{barticle}
\bauthor{\bsnm{{Vr{\v{s}}nak}}, \binits{B.}},
\bauthor{\bsnm{{Warmuth}}, \binits{A.}},
\bauthor{\bsnm{{Temmer}}, \binits{M.}},
\bauthor{\bsnm{{Veronig}}, \binits{A.}},
\bauthor{\bsnm{{Magdaleni{\'c}}}, \binits{J.}},
\bauthor{\bsnm{{Hillaris}}, \binits{A.}},
\bauthor{\bsnm{{Karlick{\'y}}}, \binits{M.}}:
\byear{2006},
\batitle{{Multi-wavelength study of coronal waves associated with the CME-flare
  event of 3 November 2003}}.
\bjtitle{\aap}
\bvolume{448}(\bissue{2}),
\bfpage{739}.
\doiurl{10.1051/0004-6361:20053740}.
\adsurl{https://ui.adsabs.harvard.edu/abs/2006A&A...448..739V}.
\end{barticle}
\endbibitem

\bibitem[\protect\citeauthoryear{Webb and Howard}{2012}]{article}
\begin{barticle}
\bauthor{\bsnm{Webb}, \binits{D.}},
\bauthor{\bsnm{Howard}, \binits{T.}}:
\byear{2012},
\batitle{Coronal mass ejections: Observations}.
\bjtitle{Liv. Rev. Solar Phys.}
\bvolume{9}.
\doiurl{10.12942/lrsp-2012-3}.
\end{barticle}
\endbibitem

\bibitem[\protect\citeauthoryear{{Yashiro} et~al.}{2004}]{2004JGRA..109.7105Y}
\begin{barticle}
\bauthor{\bsnm{{Yashiro}}, \binits{S.}},
\bauthor{\bsnm{{Gopalswamy}}, \binits{N.}},
\bauthor{\bsnm{{Michalek}}, \binits{G.}},
\bauthor{\bsnm{{St. Cyr}}, \binits{O.C.}},
\bauthor{\bsnm{{Plunkett}}, \binits{S.P.}},
\bauthor{\bsnm{{Rich}}, \binits{N.B.}},
\bauthor{\bsnm{{Howard}}, \binits{R.A.}}:
\byear{2004},
\batitle{{A catalog of white light coronal mass ejections observed by the SOHO
  spacecraft}}.
\bjtitle{Geophysi. Res. (Space Phys.)}
\bvolume{109}(\bissue{A7}),
\bfpage{A07105}.
\doiurl{10.1029/2003JA010282}.
\adsurl{https://ui.adsabs.harvard.edu/abs/2004JGRA..109.7105Y}.
\end{barticle}
\endbibitem

\bibitem[\protect\citeauthoryear{{Zhang} et~al.}{2001}]{Zhang2001}
\begin{barticle}
\bauthor{\bsnm{{Zhang}}, \binits{J.}},
\bauthor{\bsnm{{Dere}}, \binits{K.P.}},
\bauthor{\bsnm{{Howard}}, \binits{R.A.}},
\bauthor{\bsnm{{Kundu}}, \binits{M.R.}},
\bauthor{\bsnm{{White}}, \binits{S.M.}}:
\byear{2001},
\batitle{{On the Temporal Relationship between Coronal Mass Ejections and
  Flares}}.
\bjtitle{\apj}
\bvolume{559}(\bissue{1}),
\bfpage{452}.
\doiurl{10.1086/322405}.
\adsurl{https://ui.adsabs.harvard.edu/abs/2001ApJ...559..452Z}.
\end{barticle}
\endbibitem

\bibitem[\protect\citeauthoryear{Zhang et~al.}{2004}]{Zhang_2004}
\begin{barticle}
\bauthor{\bsnm{Zhang}, \binits{J.}},
\bauthor{\bsnm{Dere}, \binits{K.P.}},
\bauthor{\bsnm{Howard}, \binits{R.A.}},
\bauthor{\bsnm{Vourlidas}, \binits{A.}}:
\byear{2004},
\batitle{A study of the kinematic evolution of coronal mass ejections}.
\bjtitle{Astrophys. J.}
\bvolume{604}(\bissue{1}),
\bfpage{420}.
\doiurl{10.1086/381725}.
\end{barticle}
\endbibitem

\bibitem[\protect\citeauthoryear{{Zucca} et~al.}{2014}]{Zucca2014}
\begin{barticle}
\bauthor{\bsnm{{Zucca}}, \binits{P.}},
\bauthor{\bsnm{{Pick}}, \binits{M.}},
\bauthor{\bsnm{{D{\'e}moulin}}, \binits{P.}},
\bauthor{\bsnm{{Kerdraon}}, \binits{A.}},
\bauthor{\bsnm{{Lecacheux}}, \binits{A.}},
\bauthor{\bsnm{{Gallagher}}, \binits{P.T.}}:
\byear{2014},
\batitle{{Understanding Coronal Mass Ejections and Associated Shocks in the
  Solar Corona by Merging Multiwavelength Observations}}.
\bjtitle{\apj}
\bvolume{795}(\bissue{1}),
\bfpage{68}.
\doiurl{10.1088/0004-637X/795/1/68}.
\adsurl{https://ui.adsabs.harvard.edu/abs/2014ApJ...795...68Z}.
\end{barticle}
\endbibitem

\end{thebibliography}

\IfFileExists{\jobname.bbl}{} {\typeout{}
\typeout{****************************************************}
\typeout{****************************************************}
\typeout{** Please run "bibtex \jobname" to obtain} \typeout{**
the bibliography and then re-run LaTeX} \typeout{** twice to fix
the references !}
\typeout{****************************************************}
\typeout{****************************************************}
\typeout{}}

\end{article} 

\end{document}